\documentclass[aps,prl,twocolumn,showkeys,superscriptaddress]{revtex4-1}
\usepackage{placeins}
\usepackage[artemisia]{textgreek}
\usepackage{amssymb}
\usepackage{epsfig}
\usepackage{amsmath}
\usepackage{graphicx}
\usepackage{dcolumn}
\usepackage{latexsym}
\usepackage{color}
\usepackage{epstopdf}
\usepackage{subfigure}
\usepackage{comment}
\usepackage{nicefrac}
\usepackage{blindtext}
\usepackage[ulem=normalem]{changes}
\usepackage{float}
\usepackage{soul}
\usepackage[T1]{fontenc}
\usepackage[hidelinks]{hyperref}
\hypersetup{
    colorlinks,
    linkcolor={red!50!black},
    citecolor={blue!50!black},
    urlcolor={blue!80!black}
}

\newcommand*{ \citen}[1]{%
  \begingroup
    \romannumeral-`\x % remove space at the beginning of \setcitestyle
    \setcitestyle{numbers}%
     \cite{#1}%
  \endgroup   
}

\begin{document}

\title{Molecular beam epitaxy of CuMnAs}

\author{Filip Krizek\textsuperscript{\textdagger}}
\affiliation{Institute of Physics, Czech Academy of Sciences, Cukrovarnick{\'a} 10, 162 00 Praha 6, Czech Republic.}
\altaffiliation{These authors contributed equally to this work.}

\author{Zden\v{e}k Ka\v{s}par\textsuperscript{\textdagger}}
\affiliation{Institute of Physics, Czech Academy of Sciences,  Cukrovarnick{\'a} 10, 162 00 Praha 6, Czech Republic.}
\affiliation{Faculty of Mathematics and Physics, Charles University, Ke Karlovu 3, 121 16 Prague 2, Czech Republic}
\altaffiliation{These authors contributed equally to this work.}

\author{Aliaksei Vetushka}
\affiliation{Institute of Physics, Czech Academy of Sciences,  Cukrovarnick{\'a} 10, 162 00 Praha 6, Czech Republic.}

\author{Dominik Kriegner}
\affiliation{Institute of Physics, Czech Academy of Sciences,  Cukrovarnick{\'a} 10, 162 00 Praha 6, Czech Republic.}
\affiliation{Faculty of Mathematics and Physics, Charles University, Ke Karlovu 3, 121 16 Prague 2, Czech Republic}

\author{Elisabetta M. Fiordaliso}
\affiliation{DTU Nanolab, Technical University of Denmark,
Fysikvej, Building 307, DK-2800 Kgs. Lyngby, Denmark}

\author{Jan Michalicka}
\affiliation{Central European Institute of Technology, Brno University of Technology, Purky\v{n}ova 123, 612 00, Brno, Czech Republic}

\author{Ond\v{r}ej Man}
\affiliation{Central European Institute of Technology, Brno University of Technology, Purky\v{n}ova 123, 612 00, Brno, Czech Republic}

\author{Jan Zub{\'a}\v{c}}
\affiliation{Institute of Physics, Czech Academy of Sciences,  Cukrovarnick{\'a} 10, 162 00 Praha 6, Czech Republic.}
\affiliation{Faculty of Mathematics and Physics, Charles University, Ke Karlovu 3, 121 16 Prague 2, Czech Republic}

\author{Martin Brajer}
\affiliation{Institute of Physics, Czech Academy of Sciences,  Cukrovarnick{\'a} 10, 162 00 Praha 6, Czech Republic.}
\affiliation{Faculty of Mathematics and Physics, Charles University, Ke Karlovu 3, 121 16 Prague 2, Czech Republic}

\author{Victoria A. Hills}
\affiliation{School of Physics and Astronomy, University of Nottingham, Nottingham NG7 2RD, United Kingdom}

\author{Kevin W. Edmonds}
\affiliation{School of Physics and Astronomy, University of Nottingham, Nottingham NG7 2RD, United Kingdom}

\author{Peter Wadley}
\affiliation{School of Physics and Astronomy, University of Nottingham, Nottingham NG7 2RD, United Kingdom}

\author{Richard P. Campion}
\affiliation{School of Physics and Astronomy, University of Nottingham, Nottingham NG7 2RD, United Kingdom}

\author{Kamil Olejn{\'i}k}
\affiliation{Institute of Physics, Czech Academy of Sciences,  Cukrovarnick{\'a} 10, 162 00 Praha 6, Czech Republic.}

\author{Tom{\'a}\v{s} Jungwirth}
\affiliation{Institute of Physics, Czech Academy of Sciences,  Cukrovarnick{\'a} 10, 162 00 Praha 6, Czech Republic.}
\affiliation{School of Physics and Astronomy, University of Nottingham, Nottingham NG7 2RD, United Kingdom}

\author{V{\'i}t Nov{\'a}k}
\email{novakvit@fzu.cz}
\affiliation{Institute of Physics, Czech Academy of Sciences,  Cukrovarnick{\'a} 10, 162 00 Praha 6, Czech Republic.}

\date{\today}

%%%%%%%%%%%%%%% Abstract %%%%%%%%%%%%%%
\begin{abstract}

We present a detailed study of the growth of the tetragonal polymorph of antiferromagnetic CuMnAs by the molecular beam epitaxy technique. We explore the parameter space of growth conditions and their effect on the microstructural and transport properties of the material. We identify its typical structural defects and compare the properties of epitaxial CuMnAs layers grown on GaP, GaAs and Si substrates. Finally, we investigate the correlation between the crystalline quality of CuMnAs and its performance in terms of electrically induced resistance switching. 

\end{abstract}
% \keywords{CuMnAs, Antiferromagnetism, ....}
\maketitle

%%%%%%%%%%%%%%% Introduction %%%%%%%%%%%%%%

\section{Introduction}

The tetragonal phase of the antiferromagnetic CuMnAs \cite{wadley2013tetragonal} attracted significant attention due to the possibility to control its N\'eel vector orientation by electrical current. It was predicted for a family of collinear antiferromagnets with non-centrosymmetric spin-sublattices that a uniform electrical current induces an effective magnetic field with sign alternating between the magnetic sublattices, resulting in  reorientation of the magnetic moments \cite{vzelezny2014relativistic}. So far, current-induced N\'eel vector switching has been experimentally demonstrated in two prototypical examples of this materials family: in CuMnAs \cite{wadley2016electrical} and Mn$_{2}$Au \cite{bodnar2018writing}. Moreover, further investigation of CuMnAs devices has recently uncovered an additional mechanism of high-resistive analog switching \citen{kavspar2019high} ascribed to antiferromagnetic domain fragmentation generated by electrical and optical pulses of lengths ranging from microseconds to femtoseconds. Combined with the earlier observations of current-induced N\'eel vector switching, this makes the material potentially suitable for developing memory, opto-electronic and neuromorphic devices. The complexity of the switching mechanisms in CuMnAs and the broad range of considered device concepts call for a comprehensive materials growth and structural characterization study which is presented in this paper.     

A favorable property of CuMnAs is the high degree of structural compatibility to common semiconductor materials, specifically to GaAs, GaP and Si. This makes it possible to employ molecular beam epitaxy (MBE) as the growth technique and to take full advantage of its precise control over the growth parameters. In this work we focus on the investigation of the growth conditions and  material characterization of thin films of CuMnAs. We provide results from atomic force microscopy (AFM), superconducting quantum interference device (SQUID) magnetometry, X-ray diffraction (XRD) and transport measurement techniques, and use them to identify the optimal growth conditions of stoichiometric CuMnAs. Using scanning transmission electron microscopy (STEM) we also identify two dominant types of crystallographic defects. Finally, we relate the resistive switching performance of CuMnAs microdevices to the studied growth conditions and crystalline quality of the films.

%%%%%%%%%%%%%%% Results and Discussion %%%%%%%%%%%%%%
\section{Growth on G\lowercase{a}P}
\begin{figure}[hb!]
\vspace{0.2cm}
\includegraphics[scale=0.24]{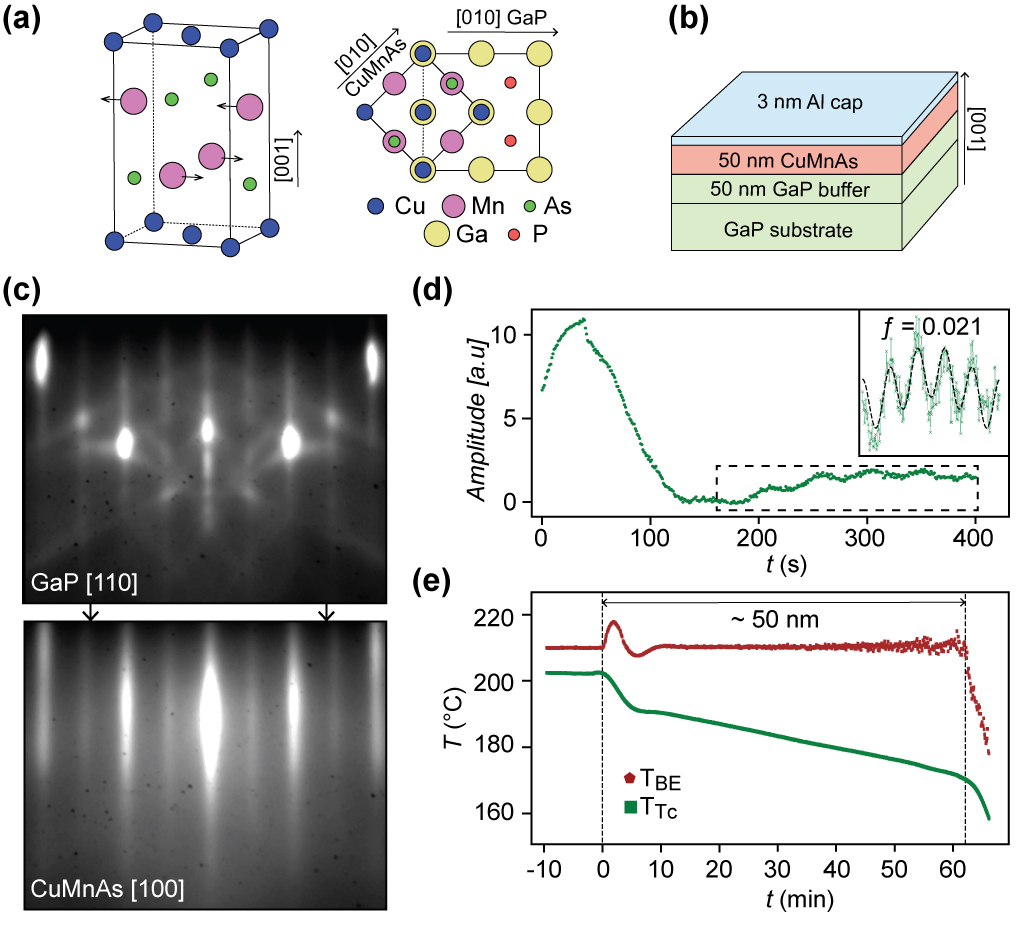}
\vspace{-0.4cm}
\caption{(a) Model of tetragonal CuMnAs unit cell, where the black arrows show the expected orientation of magnetic moments (left panel). Model of CuMnAs matching the GaP unit cell, viewed from [001] direction (right panel). (b) Sketch of the cross-sectional structure of our standard samples. (c) Typical RHEED image of 2x reconstruction of (001) GaP buffer surface taken along the [110] crystal axis (top panel). Corresponding 2x reconstruction of CuMnAs taken under the same direction after 50 nm of growth (bottom panel). (d) RHEED oscillations acquired at the beginning of CuMnAs growth, with growth rate equal to 0.021 unit cell per second. (e) Time dependent evolution of growth temperature $T_{BE}$ measured and regulated from the position of the optical absorbtion edge of the GaP substrate (red curve). Corresponding growth temperature $T_{Tc}$ measured by a thermocouple at the substrate holder (green curve).}
\label{fig1}
\end{figure}

The crystal structure of tetragonal CuMnAs \cite{wadley2013tetragonal} is shown in Fig.~\ref{fig1}(a). Antiferromagnetically ordered magnetic moments are arranged in two inversion-partner sublattices \cite{wadley2015antiferromagnetic}. Successful epitaxy of CuMnAs is achievable on GaP, GaAs and Si (001) substrates, with the CuMnAs thin film growing under 45$^{\circ}$ in-plane rotation \cite{wadley2013tetragonal}, i.e. the <100> direction of CuMnAs aligns with the <110> direction of the substrate. From the three substrate materials, GaP can be expected to yield results superior to both GaAs with large lattice mismatch and Si with non-polar surface and 1/4 unit-cell surface steps.

Our typical sample structure is shown in Fig.~\ref{fig1}(b). The buffer layer improves the quality of the GaP-CuMnAs interface and the Al capping layer serves as a protection against oxidation at ambient conditions. For our growth conditions (more details are given in Methods \cite{suppinfo}), the (2x4) GaP surface reconstruction transforms to the (2x2) CuMnAs reconstruction within the first 1-2 nm and then remains stable during further growth. The transition is shown in Fig.~\ref{fig1}(c) for the GaP buffer layer viewed along [110] direction and 50 nm thick CuMnAs viewed along the same azimuth (<100> direction of CuMnAs). After the transition period, weak reflection high-energy electron diffraction (RHEED) oscillations appear, as shown in Fig.~\ref{fig1}(d). The comparison of the layer thickness measured by X-ray reflectivity and/or atomically resolved STEM imaging shows that one RHEED oscillation period (here $\sim$ 47 s) corresponds to the growth of one full CuMnAs unit cell, in contrast to nonmagnetic zinc-blende materials with one period per one monolayer.

%%%%%%%%%%%% Fig. 1 %%%%%%%%%%%%%%%%%%%%%

A reliable substrate temperature readout is crucial for a reproducible CuMnAs growth. For this reason we measure the real-time optical absorption spectra of the substrate and extract the substrate temperature $T_{BE}$ from the position of the absorption edge during growth \citen{weilmeier1991new}. The true substrate temperature measured this way significantly deviates from the usual thermocouple, whose reading $T_{Tc}$ is delayed and more strongly linked to the substrate heater than to the substrate. If a real-time proportional-integral-derivative (PID) controller is used to maintain $T_{BE}$ constant, both heater power and $T_{Tc}$ decrease during growth with almost linear trend, as shown in  Fig.~\ref{fig1}(e). With our setup the absorption edge could be reliably resolved up to approximately 50 nm of growth. During longer growths, we extrapolate the linear decrease of the heater power to maintain a stable temperature.

%%%%%%%%%%%% Fig. 2 %%%%%%%%%%%%%%%%%%%%%
\begin{figure}[hb!]
\vspace{0.2cm}
\includegraphics[scale=0.25]{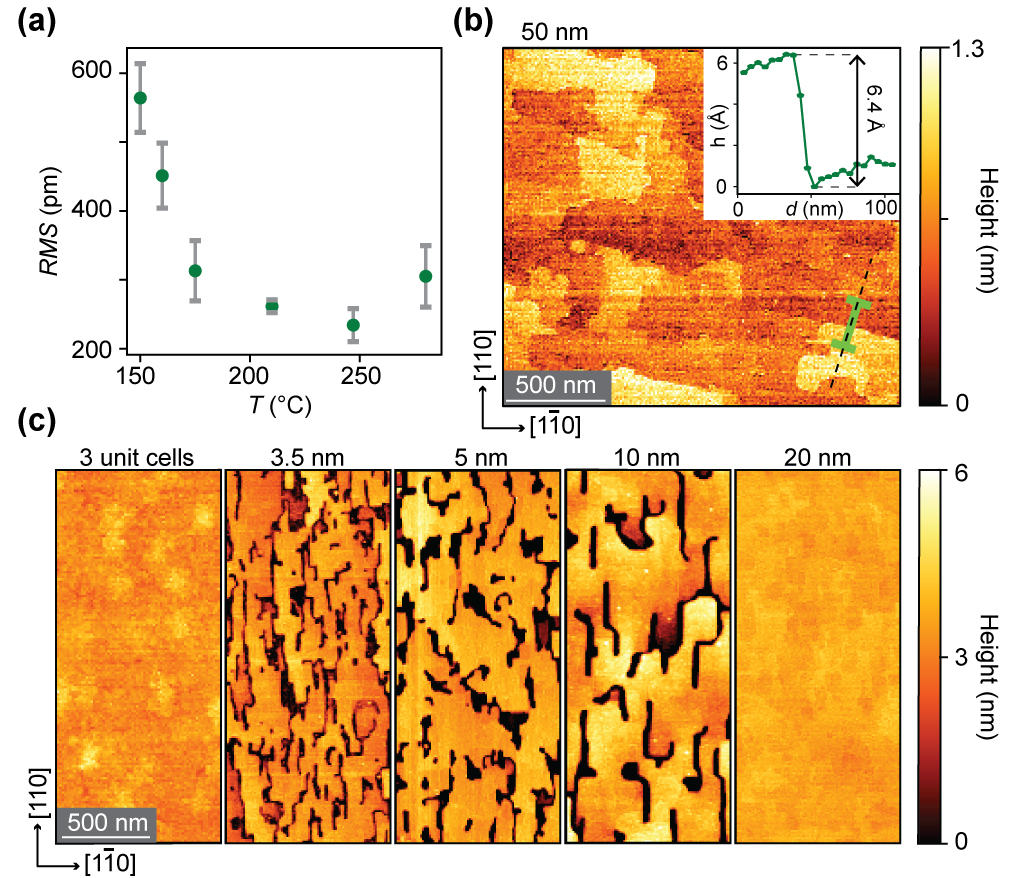}
\vspace{-0.4cm}
\caption{ (a) Dependency of surface roughness on growth temperature, for 50 nm thick CuMnAs samples grown on GaP substrate. (b) AFM micrograph of the surface of 50 nm thick CuMnAs grown at $T_{BE}=210^{\circ}$C. The height profile of a line cut through one of the terraces (green line) is shown in the inset. (c) AFM micrographs of CuMnAs layers grown on GaP with thickness varying from 1.9 nm (equivalent of 3 unit cells) to 20 nm. The highlighted crystallographic directions correspond to the orientation of the substrate.} 
\label{fig2}
\end{figure}

%%%%%%%%%%%% Fig. 3 %%%%%%%%%%%%%%%%%%%%%
\begin{figure*}[htbp!]
\vspace{0.2cm}
\includegraphics[scale=0.25]{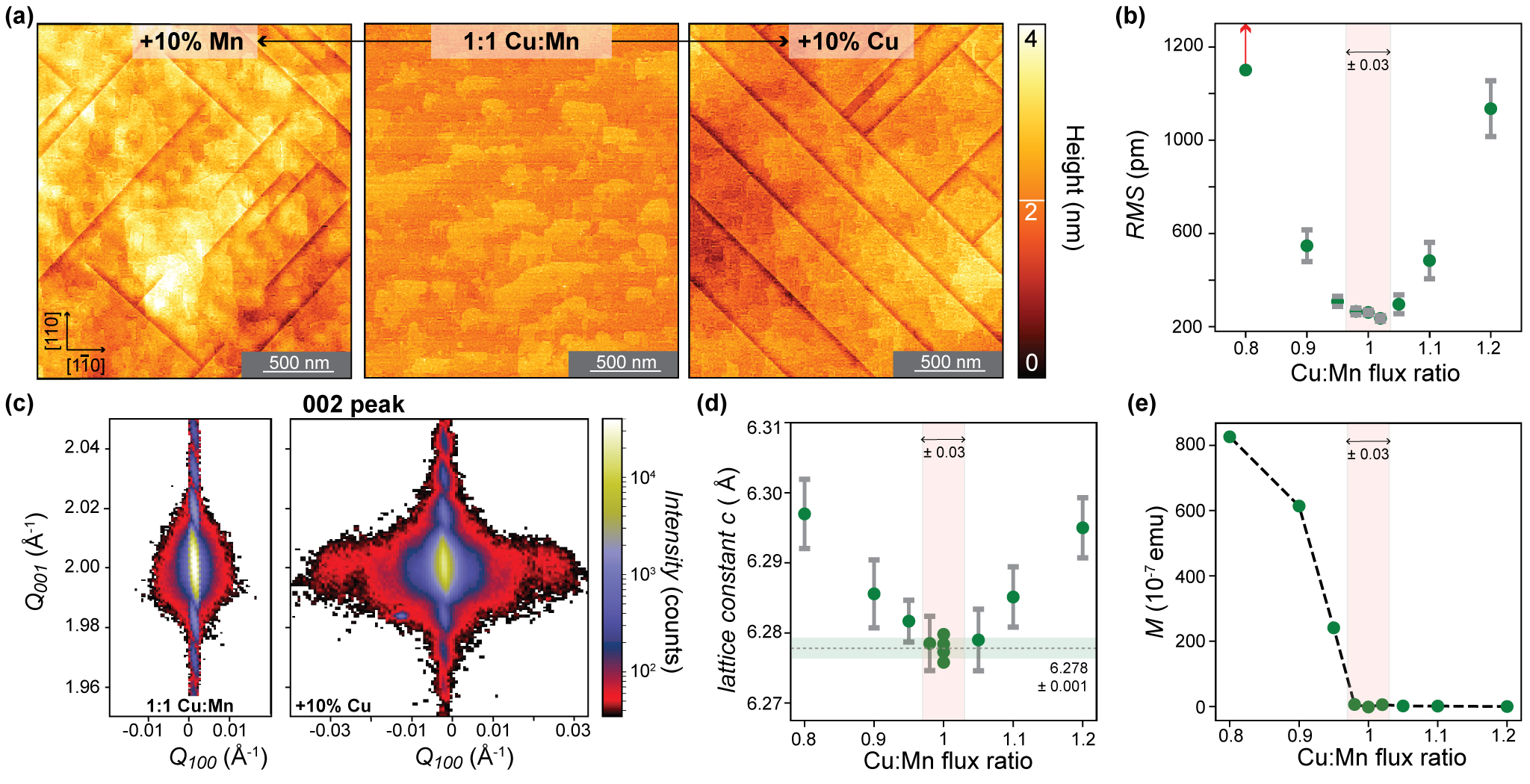}
\vspace{0.2cm}
\caption{(a) AFM micrograph of the surface of 50 nm thick Mn-rich, 1:1 stoichiometric, and Cu-rich CuMnAs samples grown on GaP. (b) Dependency of surface roughness of 50 nm thick CuMnAs layers on Cu:Mn flux ratios. The highlighted area corresponds to the typical uncertainty of the beam flux measurements by the ion gauge. (c) Reciprocal space map of 002 CuMnAs peaks for a 1:1 Cu:Mn sample (left panel) and a sample with excess copper (right panel). (d) Dependency of CuMnAs lattice constant $c$ on the Cu:Mn flux ratio. (e) Dependency of saturated magnetic moment $M$ on the Cu:Mn flux ratio. The highlighted crystallographic directions correspond to the orientation of the substrate.}
\label{fig3}
\end{figure*}

The importance of growth temperature is shown in Fig.~\ref{fig2}(a) in terms of the surface roughness of 50 nm thick CuMnAs layers grown on GaP. The measured trend reveals that optimal growth temperatures are constrained to a window from 190 to 260$^{\circ}$C. 
The surface roughness measured for samples grown within this temperature window stems mainly from terraces shown in Fig.~\ref{fig2}(b). The terraces are most frequently terminated by steps, which correspond to multiples of vertical lattice constant. This is in agreement with the unit-cell periodicity of RHEED oscillations. Both observations illustrate the tendency of the CuMnAs crystal to grow in vertical steps containing two compensated monolayers of the magnetic element.

The AFM micrographs in Fig.~\ref{fig2}(c) depict how the surface evolves for samples with varying CuMnAs thickness. The deposited material forms localized islands for the growth time equivalent to 3 unit cells. As the growth continues the islands coalesce into a percolation layer with a dense array of holes that are prevailingly elongated along the [110] direction of the GaP substrate. For even longer growth times, the density of the patterns gradually reduces until a continuous flat surface is formed. This growth regime is characteristic not only for GaP substrates, but we observe similar trends also on GaAs and Si as will be shown later in the text. The effect of the island formation in the initial stage of the growth is probably closely related to the formation of typical growth defects which will be analyzed and discussed later in the text. Under our typical growth conditions, it is possible to obtain flat and hole-free surface for thicknesses above 20 nm. 

\section{Stoichiometry}

%%%%%%%%%%%%% Fig. 4 %%%%%%%%%%%%%%%%%%%%%

\begin{figure*}[t!]
 \vspace{0.2cm}
 \includegraphics[scale=0.25]{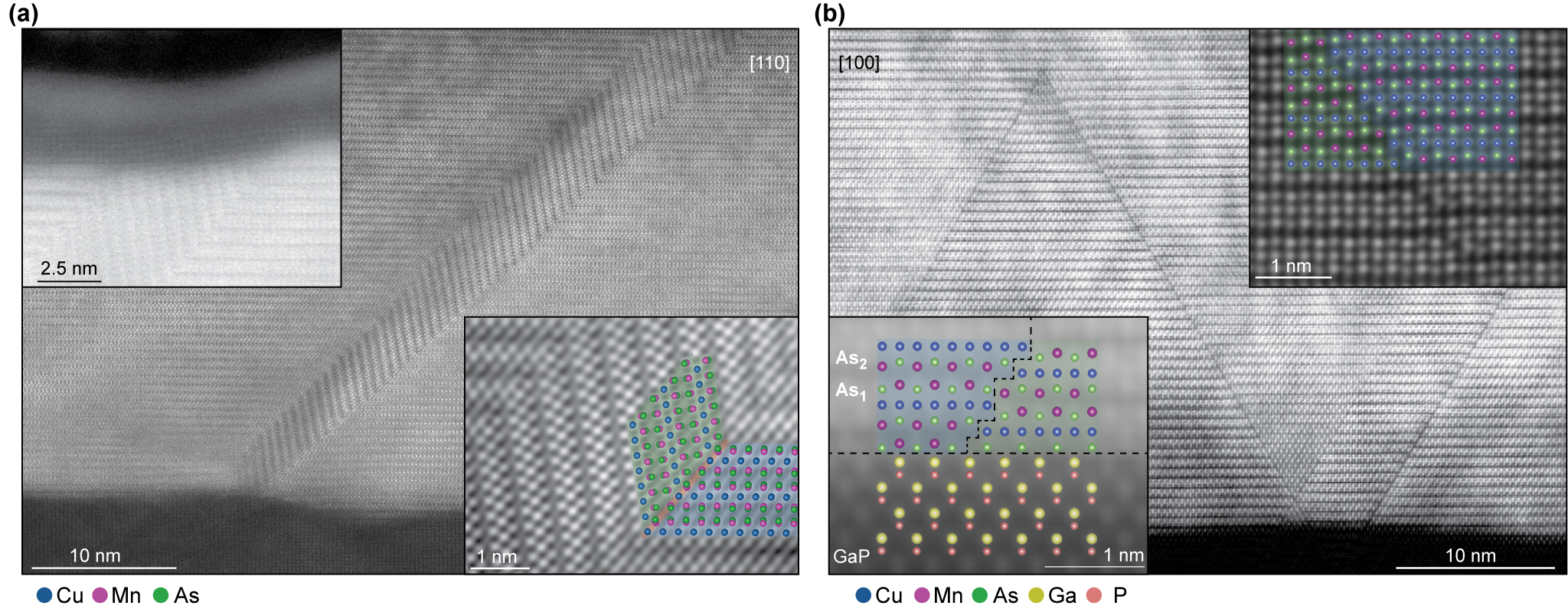}
 \vspace{0.2cm}
 \caption{(a) HAADF-STEM micrograph of a twin-like structural defect propagating throughout 50 nm thick layer of CuMnAs and viewed from [110] CuMnAs (i.e.~[100] GaP) direction. The lower inset shows a zoom in on the 
 atomic structure of the defect, overalied with expected positions of Cu - blue, Mn - purple and As - green. The uper inset then shows the detail of the top interface with the Al cap. (b) HAADF-STEM micrograph of a slip dislocations presents the same 50 nm thick layer of CuMnAs, but viewed from [100] CuMnAs (i.e.~[110] GaP) direction. The lower inset shows an atomic model overlay at the interface with the substrate (Ga - yellow, P - orange), where the first layer from the GaP substrate starts with either Mn/As layer As$_{1}$ or As$_{2}$. The upper inset shows zoom in on the atomic structure of one of the defects, with the atomic model overlay representing the expected structure.}
 \label{fig4}
 \end{figure*}

Owing to low growth temperature, the epitaxial growth of CuMnAs does not require flux overpressure of any of the elements to compensate for an incongruent re-evaporation. However, arsenic incorporation turns out to be, to a certain extent, self-controlled. We have verified that the growing crystal tolerates As flux overpressure up to a factor of 2 (relative to Cu and Mn flux intensities) without any significant effect on the crystalline quality measured by, e.g., surface morphology, defect density or electrical performance. On the other hand, these parameters sensitively depend on the ratio of the Cu and Mn fluxes. Given the limited accuracy of the independent flux calibrations and considering the unknown sticking coefficient of these elements, we have adopted the sharp minimum of the RMS surface roughness in a series of samples with varying Cu:Mn flux ratio as a sign of the 1:1:1 stoichiometry. We have verified that the composition of the corresponding layer measured by the electron-probe microanalysis (EPMA) and energy-dispersive X-ray spectroscopy (EDX) is correct within the 2\% accuracy of the methods, and we will show further in the text that the optimal stoichiometry point is consistently 
confirmed by several other parameters.

A general trend in the surface morphology as a function of the Cu:Mn flux ratio is shown in Fig.~\ref{fig3}(a). While the 1:1 Cu:Mn sample has a flat surface with terraces of the unit-cell step height, the off-stoichiometric samples exhibit characteristic line-shaped surface defects oriented along the [110] and [$\bar{1}$10] directions of CuMnAs. As shown in Fig.~\ref{fig3}(a) and (b), the density of these defects and so the RMS roughness steeply increase with the deviation from the 1:1 stoichiometry point. The defects are not related just to the surface of the layers, but they protrude through the bulk and leave a clear trace in the XRD reciprocal maps in form of a peak broadening along $Q_{100}$ axis, as shown in Fig.~\ref{fig3}(c). The broadening gradually rises for off-stoichiometric samples up to the Cu:Mn flux ratio of 1.1 and down to 0.9. Outside these limits, the broadening becomes saturated or even slightly decreases which can be attributed to precipitation of the element in excess. Let us note that the RMS roughness in the outermost points of Fig.~\ref{fig3}(b) has been measured in areas between these precipitates.

For 1:1 Cu:Mn samples grown on GaP, we extract the vertical lattice constant $c$ = 6.278 $\pm$ 0.001 \AA\, in the fully tensile-strained lattice, i.e. fully pseudomorphic growth where the lattice constant $a_{CuMnAs}$ matches $a_{GaP}/\sqrt{2}$. This value increases when the flux ratio deviates from the stoichiometry point and the lattice relaxation sets in, as shown in Fig.~\ref{fig3}(d). The lattice relaxation is likely related to the formation of micro-twin defects, associated with surface defect lines, and/or slip dislocations which will be discussed in the next section. In addition, low energy formation of Mn/Cu vacancies and/or Cu-Mn substitutions could be significant in the off-stoichiometric samples \cite{maca2019tetragonal}. In samples with more than 10\% off-stoichiometry the excess material starts forming precipitates which protrude to the surface of the samples. In case of excess Mn, these precipitates, most probably in form of MnAs, generate measurable net magnetization. This is apparent from the results of SQUID magnetometry shown in Fig.~\ref{fig3}(e). The distance between the inclusions is typically on the order of units to tens of $\mu$m, which indicates surprisingly long diffusion length of excess Cu and Mn at given growth temperatures. These results are supported by magnetic force microscopy measurements shown in Supplementary information \cite{suppinfo}.    

\section{Structural defects}

We have prepared thin lamelae from various 50 nm thick CuMnAs samples by focused ion beam (FIB) technique in order to study the atomic structure of the defects. The high-angle annular dark field - scanning transmission electron microscopy (HAADF-STEM)  analysis revealed two characteristic types of defects which were present in all studied samples. The first type is visible in the [110] projection of CuMnAs, as shown in Fig.~\ref{fig4}(a). It is of the micro-twin type and consists of a thin slab of CuMnAs crystal lattice rotated by 81.9 deg, which corresponds to the angle between (111) and ($\bar{1}\bar{1}$1) planes. The slab propagates through the whole thickness of the 50 nm CuMnAs film, mostly following the same (111) or ($\bar{1}\bar{1}$1) planes. At the surface the micro-twin slabs project as characteristic defect lines along [100] and [010] directions of CuMnAs, as shown in Fig.~\ref{fig3}(a). Based on the depth profile measured across the surface defect lines by AFM, it is possible to determine the sign of the tilt of the corresponding micro-twin beneath (see the upper inset in Fig.~\ref{fig4}(a) and the Supplementary information \cite{suppinfo}). The typical width of the micro-twins is several nanometers. As was shown in Fig.~\ref{fig3}(b), there is a strong diffraction peak broadening and its magnitude scales with the micro-twin density. It is clear that it does not directly reflect the crystalline structure of the defects. This is because the lattice spacing along the [001] direction of the twin has a different value after the rotation, and therefore does not project into the 002 CuMnAs reciprocal peaks. Yet, the STEM images in the Supplementary information \cite{suppinfo} show that there is a significant amount of strain in the CuMnAs crystal surrounding the micro-twins. Therefore, the micro-twin density is directly correlated to the amount of strain in the CuMnAs crystal, which projects as the wing-like diffraction broadening of the reciprocal peaks. Finally, it should be noted that after investigating multiple different samples, we have not found any consistent correlation between the micro-twin defect location and the local morphology of the underlying substrate (presence of surface steps, misfit dislocations, or other types of defects). The only direct correlation we have found is that the defect density increases with the deviation from the 1:1 stoichiometry point.

The second type of the characteristic defects can be seen on lamelae cut along the (100) planes of CuMnAs. Their typical structure is shown in Fig.~\ref{fig4}(b). The apparent c/2 slip dislocations form anti-phase boundaries along the (011) or (0$\bar{1}1$) planes and always start at the substrate-film interface, similarly to the micro-twin defects. They can either protrude throughout the whole thickness of the film, or annihilate in a finite depth when two such defects with opposite tilt meet. It is important to note that only a fraction of the anti-phase boundaries originate at the lattice steps on the substrate, which is otherwise typical for this type of defect \citen{wang2018designing}. Here, an additional mechanism for anti-phase boundary formation an atomically flat surface of the substrate is illustrated in the inset of Fig.~\ref{fig4}(b). While the group-V-sublattice of the substrate remains equally retained in the CuMnAs film, the tetragonal CuMnAs lattice may start either with the lower As-plane (As$_{1}$), or with the upper As-plane (As$_{2}$) rotated by 90$^{\circ}$. This corresponds to a change in the stacking of the Mn and Cu layers in the individual grains. As a result, the c/2 lattice shift anti-phase boundaries form when islands with different stacking come into contact during further growth. 

This type of defect may also be related to the formation of the holes present in the thin CuMnAs films.
As apparent from Fig.~\ref{fig2}(c), there is a transition from the growth of individual islands into a
percolation layer with holes elongated along the [110] direction of the substrate for films with thickness between 3 unit
cells and 3.5 nm. These holes can then be observed in films with thicknesses up to $\sim$ 20 nm, by which
point the vast majority have disappeared. Considering the faceted profile of the holes, a possible
explanation for their creation could be that lateral growth on these facets is locally suppressed at
certain points when several anti-phase islands meet. That would lead to a formation of faceted holes
which can remain stable until a critical adatom density for incorporation is reached at the incoherent
grain boundaries. The complexity of the hole shape would then be related to the merging of multiple
grains at once, i.e. local formation of multiple incoherent boundaries in close proximity. The
overgrowth into a homogeneous layer could be locally further suppressed by the presence of impurities or
adatom clustering within the holes \citen{fujiwara1987classification,ye2013mbe}. The hole elongation is consistent with the presence
of a diffusion anisotropy on the polar III-V surface \citen{ohta1989anisotropic}, yielding islands that are elongated in the [110]
direction of the substrate prior to coalescence. Indeed, this trend can be observed in the far left-hand image of Fig.~\ref{fig2}(c), where a long and probably anisotropic diffusion length is suggested by the large spacing, layout
and shape of the islands in films of thickness below the percolation threshold. The large spacing of the
precipitates during off-stoichiometric growth, as mentioned above, is further evidence for a long
diffusion length.

The anti-phase boundary defects were present with similar density in all samples investigated by STEM. In essence, the anti-phase boundaries appear at the bottom interface about every 30 nm. Their density is lower in the upper part of the layer due to annihilation of anti-phase boundaries with opposing tilt. We did not observe any dependency on the growth parameters within the established growth temperature window. Post growth heating experiments in the STEM have shown that the defect structure remains unchanged up to temperatures comparable to the growth temperature of optimized samples. Details are shown in the Supplemetary material \citen{suppinfo}.

%%%%%%%%%%%%% Fig. 5 %%%%%%%%%%%%%%%%%%%%%
 
\begin{figure}[t!]
 \vspace{0.2cm}
 \includegraphics[scale=0.25]{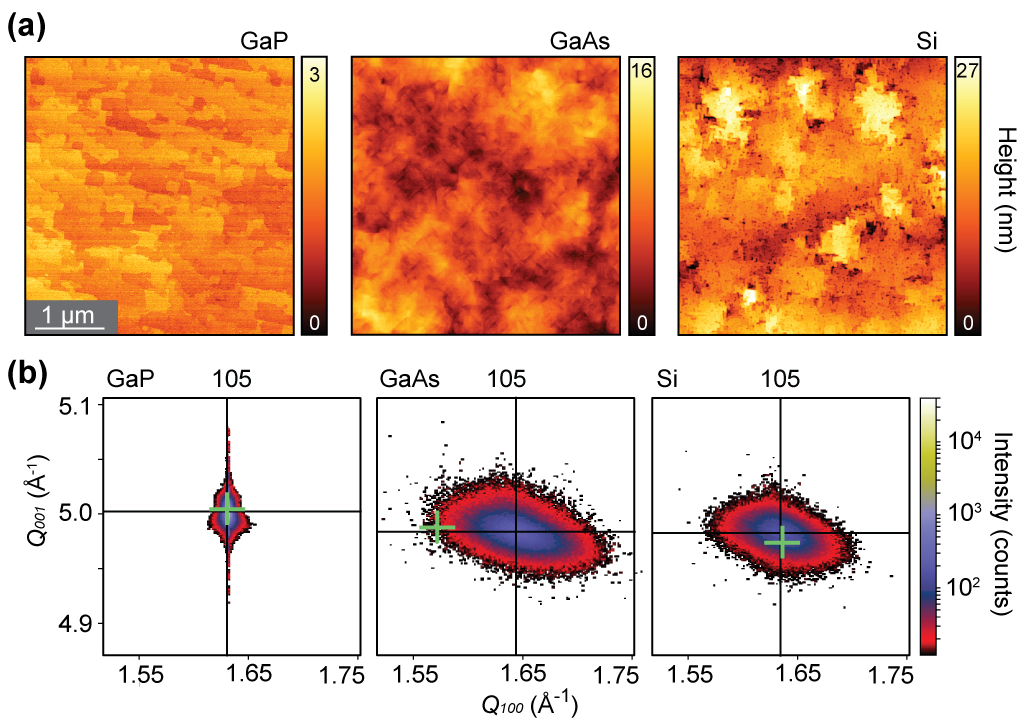}
 \vspace{0.2cm}
 \caption{(a) AFM scans of the surface of 50 nm thick CuMnAs layers grown on GaP, GaAs and Si substrates, respectively. (b) Corresponding asymmetrical reciprocal space maps. The large black cross indicates the center of the peak and the green cross the expected peak center related to the position of 204 peak of the substrate.}
 \label{fig4b}
 \end{figure}

 %%%%%%%%%%%%% Fig. 6 %%%%%%%%%%%%%%%%%%%%%
 
\begin{figure}[t!]
 \vspace{0.2cm}
 \includegraphics[scale=0.25]{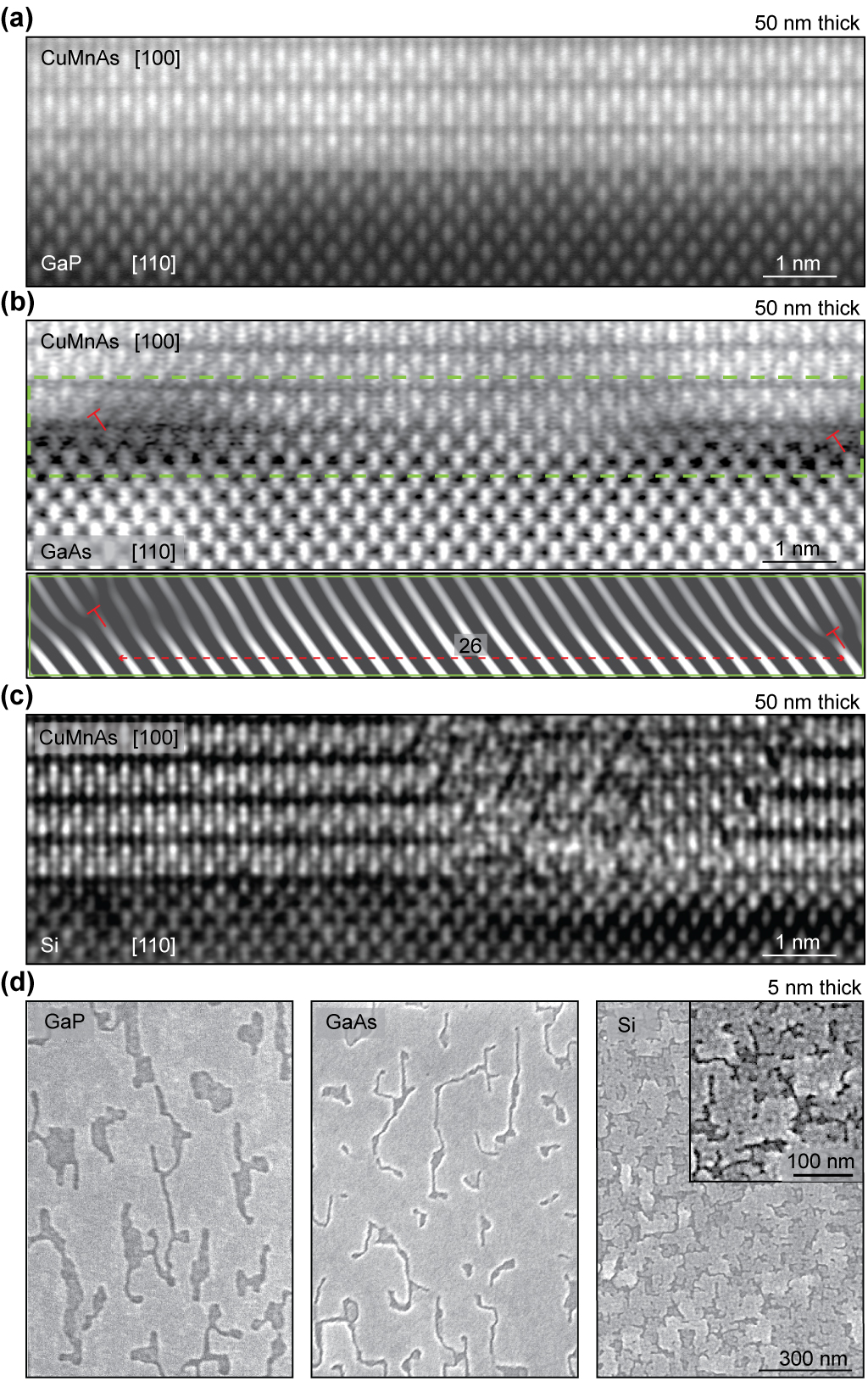}
 \vspace{0.2cm}
 \caption{(a) HAADF-STEM micrograph of the interface between CuMnAs and GaP, (b) GaAs with the area selected in green highlighting misfit dislocations (red markers) after filtering of the Fourier spectra and (c) Si. (d) SEM micrographs of 5 nm thick CuMnAs grown on GaP, GaAs and Si substrates.}
 \label{fig4b2}
 \end{figure}

 %%%%%%%%%%%% Fig. 7 %%%%%%%%%%%%%%%%%%%%%
\begin{figure}[hb!]
\vspace{0.2cm}
\includegraphics[scale=0.25]{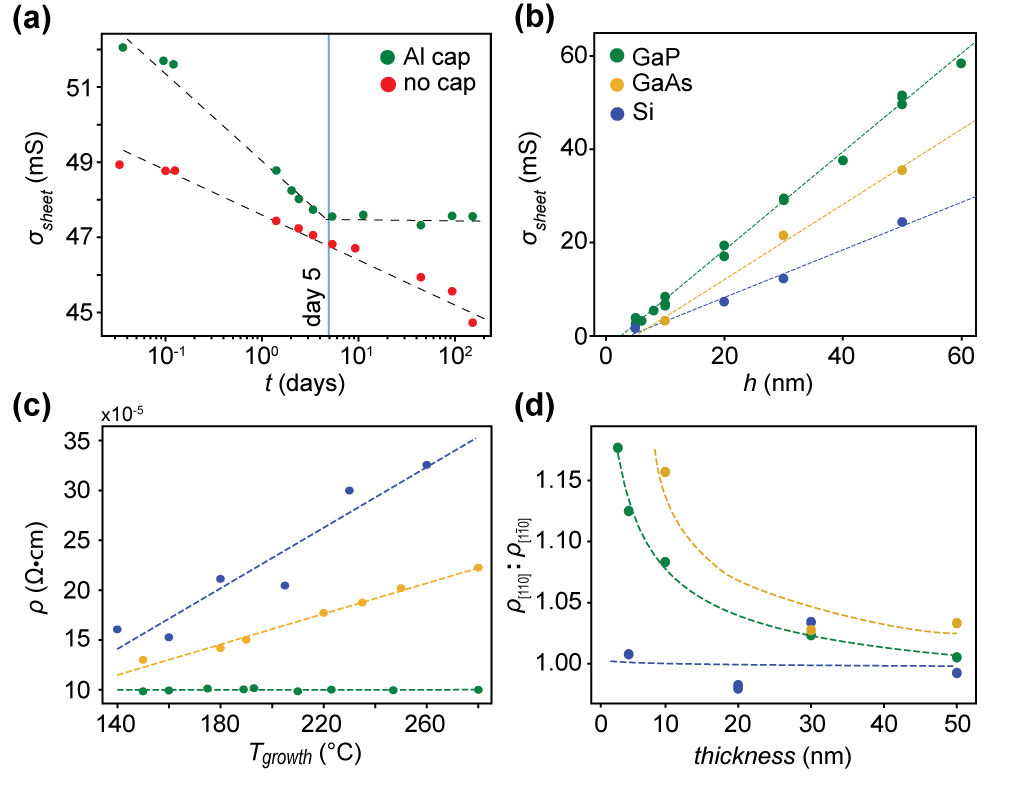}
\vspace{-0.4cm}
\caption{(a) Evolution of sheet conductance $\sigma_{sheet}$ in time for 50 nm thick CuMnAs with Al cap (red) and without capping (green) grown on GaP substrate. (b) Dependency of $\sigma_{sheet}$ on the thickness of CuMnAs films grown on GaP, GaAs and Si substrates. (c) Dependency of the resistivity $\rho$ on growth temperature for 50 nm thick films grown on GaP, GaAs and Si substrates. (d) Ratio of resistivities measured along the [110] and [0$\bar{1}$0] directions of the GaP substrate with varying thickness grown on GaP, GaAs and Si substrates. The dashed lines are guides for the eye.}
\label{fig6}
\end{figure}

%%%%%%%%%%%%% Fig. 8 %%%%%%%%%%%%%%%%%%%%%

\begin{figure*}[htbp!]
\vspace{0.2cm}
\includegraphics[scale=0.25]{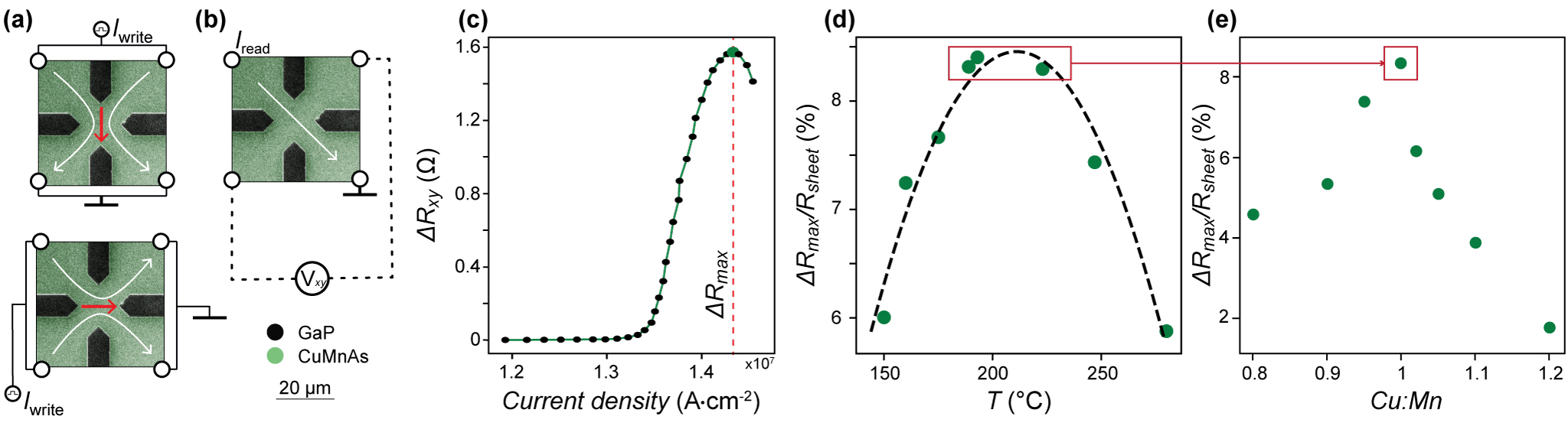}
\vspace{0.2cm}
\caption{(a) SEM micrographs of a typical cross device, where the white arrows indicate the current direction for the two orthogonal writing pulses in the top panels (net current directions are indicated by red arrow). (b) The same micrograph where the white arrow indicates the  transversal resistance readout. (c) Typical dependency of transversal resistance $\Delta$R$_{xy}$ on the pulse current density, with the maximum value $\Delta$R$_{max}$ indicated by red dashed line. (d) Dependence of the maximal difference in normalized transversal resistance $\Delta$R$_{max}$/R$_{sheet}$ on the growth temperature, with parabolic fit as guide for the eye. (e) The same dependence on Cu:Mn flux ratio. The highlighted point (red box) corresponds to the average value of samples grown within the temperature window between 180 and 220$^{\circ}$C shown in (d).}
\label{fig5}
\end{figure*}

 \section{Growth on different substrates}

Epitaxial growth of CuMnAs is also possible on (001) GaAs and Si substrates and within a similar growth temperature window. However, the surface roughness is significantly higher for films grown on GaAs and Si substrates. This is shown in the AFM images in Fig.~\ref{fig4b}(a), where the color scales correspond to the difference between the lowest and highest feature of 3 nm for GaP, 16 nm for GaAs and 27 nm for Si. The vertical lattice constant, $c$, extracted from the symmetric XRD scan along (001) axis of a 50 nm thick film is 6.299 $\pm$ 0.003 \AA\, for GaAs and 6.300 $\pm$ 0.011 \AA\, for Si; let us recall that $c$ = 6.278 $\pm$ 0.001 \AA\, in the strained layer grown on GaP. 

XRD analysis of 105 and 002 reciprocal peaks shows that there is a significant amount of mosaic tilt present in CuMnAs grown on Si and GaAs, as shown in Fig.~\ref{fig4b}(b). The peak with mosaic tilt of 0.07$^{\circ}$ sits shifted by 0.04\% with respect to the position predicted from the reference 204 peak of the GaP substrate, which is a value we attribute to the fully strained case within the error of the measurement. The situation is similar for Si substrates with a shift of 0.10\% and a mosaic tilt of 0.5$^{\circ}$. As expected from the higher theoretical mismatch on GaAs, the peak is in this case shifted by 4.59\% with a mosaic tilt of 1.0$^{\circ}$. The corresponding lattice constants $a$ extracted from the asymmetrical peaks of the optimized samples are 3.853 \AA\,  for GaP, 3.822~\AA\,  for GaAs and 3.844~\AA\,  for Si. The mosaic tilt was converted into the lateral mosaic block size by an analytical model which is described in the Supplementary information \cite{suppinfo} and was adapted from Ref. \citen{holy1999high}. The block size is the largest for CuMnAs on GaP, where it exceeds 400 nm. For GaAs we extract a smaller block size of $\sim$ 40 nm, as can be also expected due to the larger mismatch. Remarkably, the extracted block size for growth on Si is only $\sim$ 30 nm, despite the low mismatch. This is because when CuMnAs is grown on As or P rich surface, it can coherently compensate 1/2 of a lattice step on the substrate; 1/4 steps are not probable due to the group V surface termination. This is different for Si, where 1/4 steps can be present. These steps can not be coherently compensated and likely lead to the formation of defects which correlates with the increased mosaicity. In addition, a generally rougher Si surface can be expected due to the impossibility of growing Si buffer layers in our MBE system. 

The CuMnAs/substrate interface was further investigated by STEM as shown in Fig.\ref{fig4b2}. In the case of GaP, the interface is pristine and the only disturbances are the micro-twins and anti-phase boundaries described above. This is different at the CuMnAs/GaAs interface which is disturbed by arrays of misfit dislocations (with a slightly varying periodicity), as can be expected due to the large substrate/film mismatch. An example is shown in Fig.~\ref{fig4b2}(b). The periodicity is around $\sim$ 26 atomic columns and the CuMnAs layer is visibly strained around the interface. In contrast to GaP, the slip dislocations do not always protrude through the layer from the bottom interface, but often terminate within the first few nanometers. This results in local straining of the layer and additional anti-phase boundaries can appear within the strain field. Varying strain contrast and bending of the atomic rows is visible throughout the whole thickness of 50 nm samples.

The CuMnAs crystal is even more disturbed when grown on a Si substrate. An example of incoherently compensated surface step is shown in the bottom panel of Fig. \ref{fig4b2}(c). The step results in a formation of multiple crossing slip dislocations. In general, while growing on Si, the crystal structure of CuMnAs films becomes very complex. The STEM analysis shows steps on the CuMnAs surface with height up to 30 nm for 50 nm thick layers, avalanche arrays of slip dislocations and a variety of other crystallographic defects. These complex structures form above incoherent interfacial defects and disturb the film throughout the whole thickness. More STEM micrographs are shown in the Supplementary information \citen{suppinfo} for all GaP, GaAs and Si substrates.

Despite the difference in lattice mismatch, CuMnAs films thinner than 20 nm tend to form layers with holes and/or isolated islands on all three substrates, as demonstrated in Fig.~\ref{fig4b2}(d) for 5 nm thick layers. The mechanism of the formation of anti-phase islands applies in all three cases. Only the island size is smaller and the density of holes significantly higher for the Si substrate. As already mentioned, this is likely due to the presence of the 1/4 lattice steps on the surface of Si and to its non-polar surface. In contrast to films on GaAs and GaP, there is no visible anisotropy and elongation of the holes in the film while grown on the non-polar surface of Si. This observation can be linked to the (an)isotropy of the electrical resistance, as will be shown in the next section.

\section{Electronic properties }

In general, unprotected CuMnAs films tend to degrade at ambient conditions due to surface oxidation, which affects the electrical properties. This is illustrated in Fig.\ref{fig6}(a), where the sheet conductance $\sigma_{sheet}$ and its time evolution are shown for samples of 50 nm thick CuMnAs film grown on GaP, in one case unprotected and in the other case in-situ capped with 3 nm of Al. At first, both conductivities rapidly decrease. The conductivity of the capped sample started at a higher value due to the additional conductance of the Al film. However, after approximately five days, $\sigma_{sheet}$ of the capped sample stabilizes when the Al cap oxidizes and forms a stable AlO$_x$ layer. Oxidation of the unprotected layer continues with a rate that only logarithmically slows down over a period of over 150 days.

The conductance of CuMnAs films scales linearly with thickness, as shown in Fig.\ref{fig6}(b), except for ultrathin layers below approximately 3 nm, where percolation and/or quantum conductance effects step in. Apart from that, we see an onset of the conductance anisotropy for thicknesses below 20 nm, with the conductance in the [110] direction of the GaP and GaAs substrates being systematically larger than that in the [1$\bar{1}$0] direction, see Fig.\ref{fig6}(d). This is in line with the formation of the elongated islands and holes in thin films on GaP and GaAs, shown and discussed in previous sections. Let us note that no pronounced conductance anisotropy has been found in samples grown on Si, again in accordance with the absence of substrate surface polarity. We note that the anisotropy of conductance is averaged out in the value measured in the Van der Pauw geometry.

The resistivity of stoichiometric CuMnAs films grown on GaP substrates is $\rho=1.0\times10^{-4}$ $\Omega$.cm and this value is virtually independent of the growth temperature in the range from 150 to 280$^{\circ}$C, see Fig.\ref{fig6}(c). Resistivities of CuMnAs grown on GaAs or Si decrease with decreasing growth temperature, but remain higher compared to the on-GaP-grown material in the whole investigated range of temperatures. Below 140$^{\circ}$C we could perform only growth tests without the spectral temperature readout and the control of the growth temperature, resulting in a 3D growth and macroscopically rough surfaces.

The temperature dependence of the conductivity shows a clear metallic behavior on all substrates, see Supplementary information \cite{suppinfo}. The material grown on GaP showed the lowest residual resistivity at 4.2~K, 29\% of the room temperature value. This corresponds to the superior crystal quality, compared to the layers grown on GaAs and Si substrates whose residual resistivities at 4.2~K are 56\% and 85\% of the room temperature values, respectively.

In order to correlate the investigated growth parameters to the material performance in terms of current-induced resistive switching \cite{wadley2016electrical,olejnik2017antiferromagnetic,olejnik2018terahertz,kavspar2019high}, simple cross structures were prepared by optical lithography and wet etching; details of the fabrication process are given in the Supplementary information \cite{suppinfo}.
The experimental procedure and device geometry are similar to the previous work of Olejn{\'i}k et.al. \cite{olejnik2018terahertz}. Pulses of 100$\mu$s duration and varying voltage were used for generating of writing currents with density on the order of 10$^{7}$ A/cm$^{2}$. The pulses were applied in two perpendicular directions ([1$\bar{1}$0] and [110] directions of GaP, Fig.\ref{fig5}(a)), at room temperature. The transversal resistance in a 45$^{\circ}$ rotated direction (Fig.\ref{fig5}(b)) was measured 1 s after each of the writing pulses. The difference of the two transversal resistances, $\Delta$R$_{xy}$, is plotted in Fig.\ref{fig5}(c) as a function of writing current density. Here, $\Delta$R$_{xy}$ starts to rapidly grow after the pulse current exceeds a threshold value and reaches a maximum $\Delta$R$_{max}$ that is specific for the given material. We take this maximum normalized to the sheet resistance of the sample, $\Delta$R$_{max}$/R$_{sheet}$, as a material figure of merit in the switching experiments. Further increase in the pulsing voltage results in permanent change in the device response. 

Let us note that the simple cross geometry is suitable for comparison with previous studies, since the measured signal contains both the weaker spin-orbit-torque induced anisotropic magnetoresistance contribution to the switching signal \citen{vzelezny2014relativistic,wadley2016electrical,wadley2018current,grzybowski2017imaging} and the stronger increase in transversal voltage due to non-uniform change of longitudinal conductivity \citen{kavspar2019high}. The latter contribution is dominant in sample geometries optimized for measurements of longitudinal conductivity, e.g. simple bars, Wheatstone bridge geometry, etc., and is reduced in the cross geometry. We find a conversion factor of 0.4, while comparing measurements of $\Delta$R$_{max}$/R$_{sheet}$ between the cross geometry and the bar geometry presented in Ref. \citen{kavspar2019high}. The compared devices were fabricated from the same material.  

The dependence of $\Delta$R$_{max}$ on the growth temperature for the 1:1 Cu:Mn stoichiometric CuMnAs is shown in Fig.~\ref{fig5}(d). There is a clear maximum of $\Delta$R$_{max}\sim8.4$\% at growth temperatures between 190 and 230$^{\circ}$C. This correlates with the minimum surface roughness that was shown in Fig.\ref{fig2}(a). Surprisingly, there is no similar trend in the plain resistivity of the material.

The dependence of $\Delta$R$_{max}$ on the Cu:Mn ratio is shown in Fig.\ref{fig5}(e). Again, we observe a maximum $\Delta$R$_{max}$ of 8.34 $\pm$ 0.08 \% for the three 1:1 Cu:Mn samples grown within the optimal temperature window. The value of $\Delta$R$_{max}$ decreases when deviating from the 1:1 Cu:Mn stoichiometric point. This implies that the presence of the micro-twin defects does not foster the switching performance, as their density increases for off-stoichiometric samples. Moreover, the large achievable spacing between the micro-twin defects in CuMnAs grown on GaP allows us to avoid them when fabricating devices. This is hardly possible with CuMnAs grown on GaAs and Si, and the above measurements showed that $\Delta$R$_{max}$ is lower in these films, not exceeding 4.4$\%$ in case of GaAs and 4.0$\%$ in case of Si.

%%%%%%%%%%%%% Conclusion %%%%%%%%%%%%%%%%%%%%%
\section{Conclusion}

We have demonstrated that epitaxy of tetragonal CuMnAs is possible on GaP, GaAs and Si substrates. Exploring a range of growth parameters, we found a growth temperature and a material flux setting which result in a smooth surface morphology, the lowest defect density and best electrical switching performance of the CuMnAs films. We have shown that the trends of these three material properties coincide. Of the three substrates, GaP was found to result in the superior quality of the epitaxial layer.

We have identified two characteristic types of structural defects present in thin film CuMnAs. The micro-twin defects clearly correlate with the growth parameters and leave a characteristic imprint on the surface of the films and also in the reciprocal space. The occurrence of these defects can be largely suppressed in layers grown on GaP. Density of the second type of defects, the anti-phase boundaries, seems insensitive to fine adjustment of the growth conditions. This defect type is likely related to the 90-degree ambiguity in the orientation of nucleation islands of the epitaxial layer.

The observed defects may locally disturb the antiferromagnetic ordering in the material and provide potential pinning centers for magnetic domain formation \cite{jourdan2007pinning, parthasarathy2019dynamics}. A detailed theoretical study and/or further improvement of the crystal quality by implementing more complex growth protocols in order to grow anti-phase boundary free material (e.g., seeding layers, migration enhancement or growth temperature gradient) are necessary to unravel the effect which these defects may have on the performance of CuMnAs based spintronic devices.

%%%%%%%%%%%%% ACKNOWLEDGEMENT %%%%%%%%%%%%%%%%%%%%%
\section{Acknowledgement} 
We thank J. Zyka, Z. \v{S}ob{\'{a}}\v{n}, V. Jurka, K. Hru\v{s}ka and P. Zich for technical support. We also thank J.Gazquez, M. Roldan and M. Varela from Oak Ridge National Laboratory for assistance with the STEM measurements. This work was supported in part by the Ministry of Education of the Czech Republic
Grants LM2018096, LM2015087 and LNSM-LNSpin, the Grant Agency of the Czech Republic Grant No.
19-28375X, the Charles University Grant GA UK No. 886317, the EU FET Open RIA Grant
No. 766566, the Engineering and Physical Sciences Research Council Grant No. EP/P019749/1). T.J. acknowledges the support from the Neuron Foundation Prize. K.O. acknowledges the support from the Neuron Foundation Impuls Grant. Part of the work was carried out with the support of CEITEC Nano Research Infrastructure (ID LM2015041, MEYS CR, 2016-2019), CEITEC Brno University of Technology.

\bibliography{ms}

%merlin.mbs apsrev4-1.bst 2010-07-25 4.21a (PWD, AO, DPC) hacked
%Control: key (0)
%Control: author (8) initials jnrlst
%Control: editor formatted (1) identically to author
%Control: production of article title (-1) disabled
%Control: page (0) single
%Control: year (1) truncated
%Control: production of eprint (0) enabled
\begin{thebibliography}{20}%
\makeatletter
\providecommand \@ifxundefined [1]{%
 \@ifx{#1\undefined}
}%
\providecommand \@ifnum [1]{%
 \ifnum #1\expandafter \@firstoftwo
 \else \expandafter \@secondoftwo
 \fi
}%
\providecommand \@ifx [1]{%
 \ifx #1\expandafter \@firstoftwo
 \else \expandafter \@secondoftwo
 \fi
}%
\providecommand \natexlab [1]{#1}%
\providecommand \enquote  [1]{``#1''}%
\providecommand \bibnamefont  [1]{#1}%
\providecommand \bibfnamefont [1]{#1}%
\providecommand \citenamefont [1]{#1}%
\providecommand \href@noop [0]{\@secondoftwo}%
\providecommand \href [0]{\begingroup \@sanitize@url \@href}%
\providecommand \@href[1]{\@@startlink{#1}\@@href}%
\providecommand \@@href[1]{\endgroup#1\@@endlink}%
\providecommand \@sanitize@url [0]{\catcode `\\12\catcode `\$12\catcode
  `\&12\catcode `\#12\catcode `\^12\catcode `\_12\catcode `\%12\relax}%
\providecommand \@@startlink[1]{}%
\providecommand \@@endlink[0]{}%
\providecommand \url  [0]{\begingroup\@sanitize@url \@url }%
\providecommand \@url [1]{\endgroup\@href {#1}{\urlprefix }}%
\providecommand \urlprefix  [0]{URL }%
\providecommand \Eprint [0]{\href }%
\providecommand \doibase [0]{http://dx.doi.org/}%
\providecommand \selectlanguage [0]{\@gobble}%
\providecommand \bibinfo  [0]{\@secondoftwo}%
\providecommand \bibfield  [0]{\@secondoftwo}%
\providecommand \translation [1]{[#1]}%
\providecommand \BibitemOpen [0]{}%
\providecommand \bibitemStop [0]{}%
\providecommand \bibitemNoStop [0]{.\EOS\space}%
\providecommand \EOS [0]{\spacefactor3000\relax}%
\providecommand \BibitemShut  [1]{\csname bibitem#1\endcsname}%
\let\auto@bib@innerbib\@empty
%</preamble>
\bibitem [{\citenamefont {Wadley}\ \emph {et~al.}(2013)\citenamefont {Wadley},
  \citenamefont {Nov{\'a}k}, \citenamefont {Campion}, \citenamefont {Rinaldi},
  \citenamefont {Mart{\'\i}}, \citenamefont {Reichlov{\'a}}, \citenamefont
  {{\v{Z}}elezn{\'y}}, \citenamefont {Gazquez}, \citenamefont {Roldan},
  \citenamefont {Varela} \emph {et~al.}}]{wadley2013tetragonal}%
  \BibitemOpen
  \bibfield  {author} {\bibinfo {author} {\bibfnamefont {P.}~\bibnamefont
  {Wadley}}, \bibinfo {author} {\bibfnamefont {V.}~\bibnamefont {Nov{\'a}k}},
  \bibinfo {author} {\bibfnamefont {R.}~\bibnamefont {Campion}}, \bibinfo
  {author} {\bibfnamefont {C.}~\bibnamefont {Rinaldi}}, \bibinfo {author}
  {\bibfnamefont {X.}~\bibnamefont {Mart{\'\i}}}, \bibinfo {author}
  {\bibfnamefont {H.}~\bibnamefont {Reichlov{\'a}}}, \bibinfo {author}
  {\bibfnamefont {J.}~\bibnamefont {{\v{Z}}elezn{\'y}}}, \bibinfo {author}
  {\bibfnamefont {J.}~\bibnamefont {Gazquez}}, \bibinfo {author} {\bibfnamefont
  {M.}~\bibnamefont {Roldan}}, \bibinfo {author} {\bibfnamefont
  {M.}~\bibnamefont {Varela}},  \emph {et~al.},\ }\href@noop {} {\bibfield
  {journal} {\bibinfo  {journal} {Nature communications}\ }\textbf {\bibinfo
  {volume} {4}},\ \bibinfo {pages} {2322} (\bibinfo {year} {2013})}\BibitemShut
  {NoStop}%
\bibitem [{\citenamefont {{\v{Z}}elezn{\'y}}\ \emph {et~al.}(2014)\citenamefont
  {{\v{Z}}elezn{\'y}}, \citenamefont {Gao}, \citenamefont {V{\'y}born{\'y}},
  \citenamefont {Zemen}, \citenamefont {Ma{\v{s}}ek}, \citenamefont {Manchon},
  \citenamefont {Wunderlich}, \citenamefont {Sinova},\ and\ \citenamefont
  {Jungwirth}}]{vzelezny2014relativistic}%
  \BibitemOpen
  \bibfield  {author} {\bibinfo {author} {\bibfnamefont {J.}~\bibnamefont
  {{\v{Z}}elezn{\'y}}}, \bibinfo {author} {\bibfnamefont {H.}~\bibnamefont
  {Gao}}, \bibinfo {author} {\bibfnamefont {K.}~\bibnamefont
  {V{\'y}born{\'y}}}, \bibinfo {author} {\bibfnamefont {J.}~\bibnamefont
  {Zemen}}, \bibinfo {author} {\bibfnamefont {J.}~\bibnamefont {Ma{\v{s}}ek}},
  \bibinfo {author} {\bibfnamefont {A.}~\bibnamefont {Manchon}}, \bibinfo
  {author} {\bibfnamefont {J.}~\bibnamefont {Wunderlich}}, \bibinfo {author}
  {\bibfnamefont {J.}~\bibnamefont {Sinova}}, \ and\ \bibinfo {author}
  {\bibfnamefont {T.}~\bibnamefont {Jungwirth}},\ }\href@noop {} {\bibfield
  {journal} {\bibinfo  {journal} {Physical review letters}\ }\textbf {\bibinfo
  {volume} {113}},\ \bibinfo {pages} {157201} (\bibinfo {year}
  {2014})}\BibitemShut {NoStop}%
\bibitem [{\citenamefont {Wadley}\ \emph {et~al.}(2016)\citenamefont {Wadley},
  \citenamefont {Howells}, \citenamefont {{\v{Z}}elezn{\'y}}, \citenamefont
  {Andrews}, \citenamefont {Hills}, \citenamefont {Campion}, \citenamefont
  {Nov{\'a}k}, \citenamefont {Olejn{\'\i}k}, \citenamefont {Maccherozzi},
  \citenamefont {Dhesi} \emph {et~al.}}]{wadley2016electrical}%
  \BibitemOpen
  \bibfield  {author} {\bibinfo {author} {\bibfnamefont {P.}~\bibnamefont
  {Wadley}}, \bibinfo {author} {\bibfnamefont {B.}~\bibnamefont {Howells}},
  \bibinfo {author} {\bibfnamefont {J.}~\bibnamefont {{\v{Z}}elezn{\'y}}},
  \bibinfo {author} {\bibfnamefont {C.}~\bibnamefont {Andrews}}, \bibinfo
  {author} {\bibfnamefont {V.}~\bibnamefont {Hills}}, \bibinfo {author}
  {\bibfnamefont {R.~P.}\ \bibnamefont {Campion}}, \bibinfo {author}
  {\bibfnamefont {V.}~\bibnamefont {Nov{\'a}k}}, \bibinfo {author}
  {\bibfnamefont {K.}~\bibnamefont {Olejn{\'\i}k}}, \bibinfo {author}
  {\bibfnamefont {F.}~\bibnamefont {Maccherozzi}}, \bibinfo {author}
  {\bibfnamefont {S.}~\bibnamefont {Dhesi}},  \emph {et~al.},\ }\href@noop {}
  {\bibfield  {journal} {\bibinfo  {journal} {Science}\ }\textbf {\bibinfo
  {volume} {351}},\ \bibinfo {pages} {587} (\bibinfo {year}
  {2016})}\BibitemShut {NoStop}%
\bibitem [{\citenamefont {Bodnar}\ \emph {et~al.}(2018)\citenamefont {Bodnar},
  \citenamefont {{\v{S}}mejkal}, \citenamefont {Turek}, \citenamefont
  {Jungwirth}, \citenamefont {Gomonay}, \citenamefont {Sinova}, \citenamefont
  {Sapozhnik}, \citenamefont {Elmers}, \citenamefont {Kl{\"a}ui},\ and\
  \citenamefont {Jourdan}}]{bodnar2018writing}%
  \BibitemOpen
  \bibfield  {author} {\bibinfo {author} {\bibfnamefont {S.~Y.}\ \bibnamefont
  {Bodnar}}, \bibinfo {author} {\bibfnamefont {L.}~\bibnamefont
  {{\v{S}}mejkal}}, \bibinfo {author} {\bibfnamefont {I.}~\bibnamefont
  {Turek}}, \bibinfo {author} {\bibfnamefont {T.}~\bibnamefont {Jungwirth}},
  \bibinfo {author} {\bibfnamefont {O.}~\bibnamefont {Gomonay}}, \bibinfo
  {author} {\bibfnamefont {J.}~\bibnamefont {Sinova}}, \bibinfo {author}
  {\bibfnamefont {A.}~\bibnamefont {Sapozhnik}}, \bibinfo {author}
  {\bibfnamefont {H.-J.}\ \bibnamefont {Elmers}}, \bibinfo {author}
  {\bibfnamefont {M.}~\bibnamefont {Kl{\"a}ui}}, \ and\ \bibinfo {author}
  {\bibfnamefont {M.}~\bibnamefont {Jourdan}},\ }\href@noop {} {\bibfield
  {journal} {\bibinfo  {journal} {Nature communications}\ }\textbf {\bibinfo
  {volume} {9}},\ \bibinfo {pages} {348} (\bibinfo {year} {2018})}\BibitemShut
  {NoStop}%
\bibitem [{\citenamefont {Ka{\v{s}}par}\ \emph {et~al.}(2019)\citenamefont
  {Ka{\v{s}}par}, \citenamefont {Sur{\'y}nek}, \citenamefont {Zub{\'a}{\v{c}}},
  \citenamefont {Krizek}, \citenamefont {Nov{\'a}k}, \citenamefont {Campion},
  \citenamefont {W{\"o}rnle}, \citenamefont {Gambardella}, \citenamefont
  {Marti}, \citenamefont {N{\v{e}}mec} \emph {et~al.}}]{kavspar2019high}%
  \BibitemOpen
  \bibfield  {author} {\bibinfo {author} {\bibfnamefont {Z.}~\bibnamefont
  {Ka{\v{s}}par}}, \bibinfo {author} {\bibfnamefont {M.}~\bibnamefont
  {Sur{\'y}nek}}, \bibinfo {author} {\bibfnamefont {J.}~\bibnamefont
  {Zub{\'a}{\v{c}}}}, \bibinfo {author} {\bibfnamefont {F.}~\bibnamefont
  {Krizek}}, \bibinfo {author} {\bibfnamefont {V.}~\bibnamefont {Nov{\'a}k}},
  \bibinfo {author} {\bibfnamefont {R.~P.}\ \bibnamefont {Campion}}, \bibinfo
  {author} {\bibfnamefont {M.~S.}\ \bibnamefont {W{\"o}rnle}}, \bibinfo
  {author} {\bibfnamefont {P.}~\bibnamefont {Gambardella}}, \bibinfo {author}
  {\bibfnamefont {X.}~\bibnamefont {Marti}}, \bibinfo {author} {\bibfnamefont
  {P.}~\bibnamefont {N{\v{e}}mec}},  \emph {et~al.},\ }\href@noop {} {\bibfield
   {journal} {\bibinfo  {journal} {arXiv preprint arXiv:1909.09071}\ }
  (\bibinfo {year} {2019})}\BibitemShut {NoStop}%
\bibitem [{\citenamefont {Wadley}\ \emph {et~al.}(2015)\citenamefont {Wadley},
  \citenamefont {Hills}, \citenamefont {Shahedkhah}, \citenamefont {Edmonds},
  \citenamefont {Campion}, \citenamefont {Nov{\'a}k}, \citenamefont
  {Ouladdiaf}, \citenamefont {Khalyavin}, \citenamefont {Langridge},
  \citenamefont {Saidl} \emph {et~al.}}]{wadley2015antiferromagnetic}%
  \BibitemOpen
  \bibfield  {author} {\bibinfo {author} {\bibfnamefont {P.}~\bibnamefont
  {Wadley}}, \bibinfo {author} {\bibfnamefont {V.}~\bibnamefont {Hills}},
  \bibinfo {author} {\bibfnamefont {M.}~\bibnamefont {Shahedkhah}}, \bibinfo
  {author} {\bibfnamefont {K.}~\bibnamefont {Edmonds}}, \bibinfo {author}
  {\bibfnamefont {R.}~\bibnamefont {Campion}}, \bibinfo {author} {\bibfnamefont
  {V.}~\bibnamefont {Nov{\'a}k}}, \bibinfo {author} {\bibfnamefont
  {B.}~\bibnamefont {Ouladdiaf}}, \bibinfo {author} {\bibfnamefont
  {D.}~\bibnamefont {Khalyavin}}, \bibinfo {author} {\bibfnamefont
  {S.}~\bibnamefont {Langridge}}, \bibinfo {author} {\bibfnamefont
  {V.}~\bibnamefont {Saidl}},  \emph {et~al.},\ }\href@noop {} {\bibfield
  {journal} {\bibinfo  {journal} {Scientific reports}\ }\textbf {\bibinfo
  {volume} {5}},\ \bibinfo {pages} {17079} (\bibinfo {year}
  {2015})}\BibitemShut {NoStop}%
\bibitem [{sup()}]{suppinfo}%
  \BibitemOpen
  \href@noop {} {\bibinfo  {journal} {See Supplemental Material at [URL will be
  inserted by publisher] for following details: Methods; S1: Surface line
  defects and magnetic inclusions; S2: Identification of micro-twin tilt in
  AFM; S3: Strain induced by the micro-twin defects; S4: Atomic structure of
  CuMnAs grown on different substrates; S5: Effect of heating on CuMnAs; S6:
  Temperature dependence of resistance.}\ }\BibitemShut {NoStop}%
\bibitem [{\citenamefont {Weilmeier}\ \emph {et~al.}(1991)\citenamefont
  {Weilmeier}, \citenamefont {Colbow}, \citenamefont {Tiedje}, \citenamefont
  {Buuren},\ and\ \citenamefont {Xu}}]{weilmeier1991new}%
  \BibitemOpen
\bibfield  {journal} {  }\bibfield  {author} {\bibinfo {author} {\bibfnamefont
  {M.}~\bibnamefont {Weilmeier}}, \bibinfo {author} {\bibfnamefont
  {K.}~\bibnamefont {Colbow}}, \bibinfo {author} {\bibfnamefont
  {T.}~\bibnamefont {Tiedje}}, \bibinfo {author} {\bibfnamefont {T.~V.}\
  \bibnamefont {Buuren}}, \ and\ \bibinfo {author} {\bibfnamefont
  {L.}~\bibnamefont {Xu}},\ }\href@noop {} {\bibfield  {journal} {\bibinfo
  {journal} {Canadian Journal of Physics}\ }\textbf {\bibinfo {volume} {69}},\
  \bibinfo {pages} {422} (\bibinfo {year} {1991})}\BibitemShut {NoStop}%
\bibitem [{\citenamefont {M{\'a}ca}\ \emph {et~al.}(2019)\citenamefont
  {M{\'a}ca}, \citenamefont {Kudrnovsk{\'y}}, \citenamefont {Bal{\'a}{\v{z}}},
  \citenamefont {Drchal}, \citenamefont {Carva},\ and\ \citenamefont
  {Turek}}]{maca2019tetragonal}%
  \BibitemOpen
  \bibfield  {author} {\bibinfo {author} {\bibfnamefont {F.}~\bibnamefont
  {M{\'a}ca}}, \bibinfo {author} {\bibfnamefont {J.}~\bibnamefont
  {Kudrnovsk{\'y}}}, \bibinfo {author} {\bibfnamefont {P.}~\bibnamefont
  {Bal{\'a}{\v{z}}}}, \bibinfo {author} {\bibfnamefont {V.}~\bibnamefont
  {Drchal}}, \bibinfo {author} {\bibfnamefont {K.}~\bibnamefont {Carva}}, \
  and\ \bibinfo {author} {\bibfnamefont {I.}~\bibnamefont {Turek}},\
  }\href@noop {} {\bibfield  {journal} {\bibinfo  {journal} {Journal of
  Magnetism and Magnetic Materials}\ }\textbf {\bibinfo {volume} {474}},\
  \bibinfo {pages} {467} (\bibinfo {year} {2019})}\BibitemShut {NoStop}%
\bibitem [{\citenamefont {Wang}\ \emph {et~al.}(2018)\citenamefont {Wang},
  \citenamefont {Guo}, \citenamefont {Shao}, \citenamefont {Saghayezhian},
  \citenamefont {Li}, \citenamefont {Fittipaldi}, \citenamefont {Vecchione},
  \citenamefont {Siwakoti}, \citenamefont {Zhu}, \citenamefont {Zhang} \emph
  {et~al.}}]{wang2018designing}%
  \BibitemOpen
  \bibfield  {author} {\bibinfo {author} {\bibfnamefont {Z.}~\bibnamefont
  {Wang}}, \bibinfo {author} {\bibfnamefont {H.}~\bibnamefont {Guo}}, \bibinfo
  {author} {\bibfnamefont {S.}~\bibnamefont {Shao}}, \bibinfo {author}
  {\bibfnamefont {M.}~\bibnamefont {Saghayezhian}}, \bibinfo {author}
  {\bibfnamefont {J.}~\bibnamefont {Li}}, \bibinfo {author} {\bibfnamefont
  {R.}~\bibnamefont {Fittipaldi}}, \bibinfo {author} {\bibfnamefont
  {A.}~\bibnamefont {Vecchione}}, \bibinfo {author} {\bibfnamefont
  {P.}~\bibnamefont {Siwakoti}}, \bibinfo {author} {\bibfnamefont
  {Y.}~\bibnamefont {Zhu}}, \bibinfo {author} {\bibfnamefont {J.}~\bibnamefont
  {Zhang}},  \emph {et~al.},\ }\href@noop {} {\bibfield  {journal} {\bibinfo
  {journal} {Proceedings of the National Academy of Sciences}\ }\textbf
  {\bibinfo {volume} {115}},\ \bibinfo {pages} {9485} (\bibinfo {year}
  {2018})}\BibitemShut {NoStop}%
\bibitem [{\citenamefont {Fujiwara}\ \emph {et~al.}(1987)\citenamefont
  {Fujiwara}, \citenamefont {Kanamoto}, \citenamefont {Ohta}, \citenamefont
  {Tokuda},\ and\ \citenamefont {Nakayama}}]{fujiwara1987classification}%
  \BibitemOpen
  \bibfield  {author} {\bibinfo {author} {\bibfnamefont {K.}~\bibnamefont
  {Fujiwara}}, \bibinfo {author} {\bibfnamefont {K.}~\bibnamefont {Kanamoto}},
  \bibinfo {author} {\bibfnamefont {Y.}~\bibnamefont {Ohta}}, \bibinfo {author}
  {\bibfnamefont {Y.}~\bibnamefont {Tokuda}}, \ and\ \bibinfo {author}
  {\bibfnamefont {T.}~\bibnamefont {Nakayama}},\ }\href@noop {} {\bibfield
  {journal} {\bibinfo  {journal} {Journal of crystal growth}\ }\textbf
  {\bibinfo {volume} {80}},\ \bibinfo {pages} {104} (\bibinfo {year}
  {1987})}\BibitemShut {NoStop}%
\bibitem [{\citenamefont {Ye}\ \emph {et~al.}(2013)\citenamefont {Ye},
  \citenamefont {Li}, \citenamefont {Hinkey}, \citenamefont {Yang},
  \citenamefont {Mishima}, \citenamefont {Keay}, \citenamefont {Santos},\ and\
  \citenamefont {Johnson}}]{ye2013mbe}%
  \BibitemOpen
  \bibfield  {author} {\bibinfo {author} {\bibfnamefont {H.}~\bibnamefont
  {Ye}}, \bibinfo {author} {\bibfnamefont {L.}~\bibnamefont {Li}}, \bibinfo
  {author} {\bibfnamefont {R.~T.}\ \bibnamefont {Hinkey}}, \bibinfo {author}
  {\bibfnamefont {R.~Q.}\ \bibnamefont {Yang}}, \bibinfo {author}
  {\bibfnamefont {T.~D.}\ \bibnamefont {Mishima}}, \bibinfo {author}
  {\bibfnamefont {J.~C.}\ \bibnamefont {Keay}}, \bibinfo {author}
  {\bibfnamefont {M.~B.}\ \bibnamefont {Santos}}, \ and\ \bibinfo {author}
  {\bibfnamefont {M.~B.}\ \bibnamefont {Johnson}},\ }\href@noop {} {\bibfield
  {journal} {\bibinfo  {journal} {Journal of Vacuum Science \& Technology B,
  Nanotechnology and Microelectronics: Materials, Processing, Measurement, and
  Phenomena}\ }\textbf {\bibinfo {volume} {31}},\ \bibinfo {pages} {03C135}
  (\bibinfo {year} {2013})}\BibitemShut {NoStop}%
\bibitem [{\citenamefont {Ohta}\ \emph {et~al.}(1989)\citenamefont {Ohta},
  \citenamefont {Kojima},\ and\ \citenamefont
  {Nakagawa}}]{ohta1989anisotropic}%
  \BibitemOpen
  \bibfield  {author} {\bibinfo {author} {\bibfnamefont {K.}~\bibnamefont
  {Ohta}}, \bibinfo {author} {\bibfnamefont {T.}~\bibnamefont {Kojima}}, \ and\
  \bibinfo {author} {\bibfnamefont {T.}~\bibnamefont {Nakagawa}},\ }\href@noop
  {} {\bibfield  {journal} {\bibinfo  {journal} {Journal of Crystal Growth}\
  }\textbf {\bibinfo {volume} {95}},\ \bibinfo {pages} {71} (\bibinfo {year}
  {1989})}\BibitemShut {NoStop}%
\bibitem [{\citenamefont {Holy}\ \emph {et~al.}(1999)\citenamefont {Holy},
  \citenamefont {Baumbach},\ and\ \citenamefont {Pietsch}}]{holy1999high}%
  \BibitemOpen
  \bibfield  {author} {\bibinfo {author} {\bibfnamefont {V.}~\bibnamefont
  {Holy}}, \bibinfo {author} {\bibfnamefont {T.}~\bibnamefont {Baumbach}}, \
  and\ \bibinfo {author} {\bibfnamefont {U.}~\bibnamefont {Pietsch}},\
  }\href@noop {} {\emph {\bibinfo {title} {High-resolution X-ray scattering
  from thin films and multilayers}}}\ (\bibinfo  {publisher} {Springer},\
  \bibinfo {year} {1999})\BibitemShut {NoStop}%
\bibitem [{\citenamefont {Olejn{\'\i}k}\ \emph {et~al.}(2017)\citenamefont
  {Olejn{\'\i}k}, \citenamefont {Schuler}, \citenamefont {Mart{\'\i}},
  \citenamefont {Nov{\'a}k}, \citenamefont {Ka{\v{s}}par}, \citenamefont
  {Wadley}, \citenamefont {Campion}, \citenamefont {Edmonds}, \citenamefont
  {Gallagher}, \citenamefont {Garc{\'e}s} \emph
  {et~al.}}]{olejnik2017antiferromagnetic}%
  \BibitemOpen
  \bibfield  {author} {\bibinfo {author} {\bibfnamefont {K.}~\bibnamefont
  {Olejn{\'\i}k}}, \bibinfo {author} {\bibfnamefont {V.}~\bibnamefont
  {Schuler}}, \bibinfo {author} {\bibfnamefont {X.}~\bibnamefont {Mart{\'\i}}},
  \bibinfo {author} {\bibfnamefont {V.}~\bibnamefont {Nov{\'a}k}}, \bibinfo
  {author} {\bibfnamefont {Z.}~\bibnamefont {Ka{\v{s}}par}}, \bibinfo {author}
  {\bibfnamefont {P.}~\bibnamefont {Wadley}}, \bibinfo {author} {\bibfnamefont
  {R.~P.}\ \bibnamefont {Campion}}, \bibinfo {author} {\bibfnamefont {K.~W.}\
  \bibnamefont {Edmonds}}, \bibinfo {author} {\bibfnamefont {B.~L.}\
  \bibnamefont {Gallagher}}, \bibinfo {author} {\bibfnamefont {J.}~\bibnamefont
  {Garc{\'e}s}},  \emph {et~al.},\ }\href@noop {} {\bibfield  {journal}
  {\bibinfo  {journal} {Nature communications}\ }\textbf {\bibinfo {volume}
  {8}},\ \bibinfo {pages} {15434} (\bibinfo {year} {2017})}\BibitemShut
  {NoStop}%
\bibitem [{\citenamefont {Olejn{\'\i}k}\ \emph {et~al.}(2018)\citenamefont
  {Olejn{\'\i}k}, \citenamefont {Seifert}, \citenamefont {Ka{\v{s}}par},
  \citenamefont {Nov{\'a}k}, \citenamefont {Wadley}, \citenamefont {Campion},
  \citenamefont {Baumgartner}, \citenamefont {Gambardella}, \citenamefont
  {N{\v{e}}mec}, \citenamefont {Wunderlich} \emph
  {et~al.}}]{olejnik2018terahertz}%
  \BibitemOpen
  \bibfield  {author} {\bibinfo {author} {\bibfnamefont {K.}~\bibnamefont
  {Olejn{\'\i}k}}, \bibinfo {author} {\bibfnamefont {T.}~\bibnamefont
  {Seifert}}, \bibinfo {author} {\bibfnamefont {Z.}~\bibnamefont
  {Ka{\v{s}}par}}, \bibinfo {author} {\bibfnamefont {V.}~\bibnamefont
  {Nov{\'a}k}}, \bibinfo {author} {\bibfnamefont {P.}~\bibnamefont {Wadley}},
  \bibinfo {author} {\bibfnamefont {R.~P.}\ \bibnamefont {Campion}}, \bibinfo
  {author} {\bibfnamefont {M.}~\bibnamefont {Baumgartner}}, \bibinfo {author}
  {\bibfnamefont {P.}~\bibnamefont {Gambardella}}, \bibinfo {author}
  {\bibfnamefont {P.}~\bibnamefont {N{\v{e}}mec}}, \bibinfo {author}
  {\bibfnamefont {J.}~\bibnamefont {Wunderlich}},  \emph {et~al.},\ }\href@noop
  {} {\bibfield  {journal} {\bibinfo  {journal} {Science advances}\ }\textbf
  {\bibinfo {volume} {4}},\ \bibinfo {pages} {eaar3566} (\bibinfo {year}
  {2018})}\BibitemShut {NoStop}%
\bibitem [{\citenamefont {Wadley}\ \emph {et~al.}(2018)\citenamefont {Wadley},
  \citenamefont {Reimers}, \citenamefont {Grzybowski}, \citenamefont {Andrews},
  \citenamefont {Wang}, \citenamefont {Chauhan}, \citenamefont {Gallagher},
  \citenamefont {Campion}, \citenamefont {Edmonds}, \citenamefont {Dhesi} \emph
  {et~al.}}]{wadley2018current}%
  \BibitemOpen
  \bibfield  {author} {\bibinfo {author} {\bibfnamefont {P.}~\bibnamefont
  {Wadley}}, \bibinfo {author} {\bibfnamefont {S.}~\bibnamefont {Reimers}},
  \bibinfo {author} {\bibfnamefont {M.~J.}\ \bibnamefont {Grzybowski}},
  \bibinfo {author} {\bibfnamefont {C.}~\bibnamefont {Andrews}}, \bibinfo
  {author} {\bibfnamefont {M.}~\bibnamefont {Wang}}, \bibinfo {author}
  {\bibfnamefont {J.~S.}\ \bibnamefont {Chauhan}}, \bibinfo {author}
  {\bibfnamefont {B.~L.}\ \bibnamefont {Gallagher}}, \bibinfo {author}
  {\bibfnamefont {R.~P.}\ \bibnamefont {Campion}}, \bibinfo {author}
  {\bibfnamefont {K.~W.}\ \bibnamefont {Edmonds}}, \bibinfo {author}
  {\bibfnamefont {S.~S.}\ \bibnamefont {Dhesi}},  \emph {et~al.},\ }\href@noop
  {} {\bibfield  {journal} {\bibinfo  {journal} {Nature nanotechnology}\
  }\textbf {\bibinfo {volume} {13}},\ \bibinfo {pages} {362} (\bibinfo {year}
  {2018})}\BibitemShut {NoStop}%
\bibitem [{\citenamefont {Grzybowski}\ \emph {et~al.}(2017)\citenamefont
  {Grzybowski}, \citenamefont {Wadley}, \citenamefont {Edmonds}, \citenamefont
  {Beardsley}, \citenamefont {Hills}, \citenamefont {Campion}, \citenamefont
  {Gallagher}, \citenamefont {Chauhan}, \citenamefont {Novak}, \citenamefont
  {Jungwirth} \emph {et~al.}}]{grzybowski2017imaging}%
  \BibitemOpen
  \bibfield  {author} {\bibinfo {author} {\bibfnamefont {M.}~\bibnamefont
  {Grzybowski}}, \bibinfo {author} {\bibfnamefont {P.}~\bibnamefont {Wadley}},
  \bibinfo {author} {\bibfnamefont {K.}~\bibnamefont {Edmonds}}, \bibinfo
  {author} {\bibfnamefont {R.}~\bibnamefont {Beardsley}}, \bibinfo {author}
  {\bibfnamefont {V.}~\bibnamefont {Hills}}, \bibinfo {author} {\bibfnamefont
  {R.}~\bibnamefont {Campion}}, \bibinfo {author} {\bibfnamefont
  {B.}~\bibnamefont {Gallagher}}, \bibinfo {author} {\bibfnamefont {J.~S.}\
  \bibnamefont {Chauhan}}, \bibinfo {author} {\bibfnamefont {V.}~\bibnamefont
  {Novak}}, \bibinfo {author} {\bibfnamefont {T.}~\bibnamefont {Jungwirth}},
  \emph {et~al.},\ }\href@noop {} {\bibfield  {journal} {\bibinfo  {journal}
  {Physical review letters}\ }\textbf {\bibinfo {volume} {118}},\ \bibinfo
  {pages} {057701} (\bibinfo {year} {2017})}\BibitemShut {NoStop}%
\bibitem [{\citenamefont {Jourdan}\ \emph {et~al.}(2007)\citenamefont
  {Jourdan}, \citenamefont {Lan{\c{c}}on},\ and\ \citenamefont
  {Marty}}]{jourdan2007pinning}%
  \BibitemOpen
  \bibfield  {author} {\bibinfo {author} {\bibfnamefont {T.}~\bibnamefont
  {Jourdan}}, \bibinfo {author} {\bibfnamefont {F.}~\bibnamefont
  {Lan{\c{c}}on}}, \ and\ \bibinfo {author} {\bibfnamefont {A.}~\bibnamefont
  {Marty}},\ }\href@noop {} {\bibfield  {journal} {\bibinfo  {journal}
  {Physical Review B}\ }\textbf {\bibinfo {volume} {75}},\ \bibinfo {pages}
  {094422} (\bibinfo {year} {2007})}\BibitemShut {NoStop}%
\bibitem [{\citenamefont {Parthasarathy}\ and\ \citenamefont
  {Rakheja}(2019)}]{parthasarathy2019dynamics}%
  \BibitemOpen
  \bibfield  {author} {\bibinfo {author} {\bibfnamefont {A.}~\bibnamefont
  {Parthasarathy}}\ and\ \bibinfo {author} {\bibfnamefont {S.}~\bibnamefont
  {Rakheja}},\ }\href@noop {} {\bibfield  {journal} {\bibinfo  {journal}
  {Physical Review Applied}\ }\textbf {\bibinfo {volume} {11}},\ \bibinfo
  {pages} {034051} (\bibinfo {year} {2019})}\BibitemShut {NoStop}%
\end{thebibliography}%


%merlin.mbs apsrev4-1.bst 2010-07-25 4.21a (PWD, AO, DPC) hacked
%Control: key (0)
%Control: author (8) initials jnrlst
%Control: editor formatted (1) identically to author
%Control: production of article title (-1) disabled
%Control: page (0) single
%Control: year (1) truncated
%Control: production of eprint (0) enabled
\begin{thebibliography}{6}%
\makeatletter
\providecommand \@ifxundefined [1]{%
 \@ifx{#1\undefined}
}%
\providecommand \@ifnum [1]{%
 \ifnum #1\expandafter \@firstoftwo
 \else \expandafter \@secondoftwo
 \fi
}%
\providecommand \@ifx [1]{%
 \ifx #1\expandafter \@firstoftwo
 \else \expandafter \@secondoftwo
 \fi
}%
\providecommand \natexlab [1]{#1}%
\providecommand \enquote  [1]{``#1''}%
\providecommand \bibnamefont  [1]{#1}%
\providecommand \bibfnamefont [1]{#1}%
\providecommand \citenamefont [1]{#1}%
\providecommand \href@noop [0]{\@secondoftwo}%
\providecommand \href [0]{\begingroup \@sanitize@url \@href}%
\providecommand \@href[1]{\@@startlink{#1}\@@href}%
\providecommand \@@href[1]{\endgroup#1\@@endlink}%
\providecommand \@sanitize@url [0]{\catcode `\\12\catcode `\$12\catcode
  `\&12\catcode `\#12\catcode `\^12\catcode `\_12\catcode `\%12\relax}%
\providecommand \@@startlink[1]{}%
\providecommand \@@endlink[0]{}%
\providecommand \url  [0]{\begingroup\@sanitize@url \@url }%
\providecommand \@url [1]{\endgroup\@href {#1}{\urlprefix }}%
\providecommand \urlprefix  [0]{URL }%
\providecommand \Eprint [0]{\href }%
\providecommand \doibase [0]{http://dx.doi.org/}%
\providecommand \selectlanguage [0]{\@gobble}%
\providecommand \bibinfo  [0]{\@secondoftwo}%
\providecommand \bibfield  [0]{\@secondoftwo}%
\providecommand \translation [1]{[#1]}%
\providecommand \BibitemOpen [0]{}%
\providecommand \bibitemStop [0]{}%
\providecommand \bibitemNoStop [0]{.\EOS\space}%
\providecommand \EOS [0]{\spacefactor3000\relax}%
\providecommand \BibitemShut  [1]{\csname bibitem#1\endcsname}%
\let\auto@bib@innerbib\@empty
%</preamble>
\bibitem [{\citenamefont {Nov{\'a}k}\ \emph {et~al.}(2007)\citenamefont
  {Nov{\'a}k}, \citenamefont {Olejn{\'\i}k}, \citenamefont {Cukr},
  \citenamefont {Smr{\v{c}}ka}, \citenamefont {Reme{\v{s}}},\ and\
  \citenamefont {Oswald}}]{novak2007substrate}%
  \BibitemOpen
  \bibfield  {author} {\bibinfo {author} {\bibfnamefont {V.}~\bibnamefont
  {Nov{\'a}k}}, \bibinfo {author} {\bibfnamefont {K.}~\bibnamefont
  {Olejn{\'\i}k}}, \bibinfo {author} {\bibfnamefont {M.}~\bibnamefont {Cukr}},
  \bibinfo {author} {\bibfnamefont {L.}~\bibnamefont {Smr{\v{c}}ka}}, \bibinfo
  {author} {\bibfnamefont {Z.}~\bibnamefont {Reme{\v{s}}}}, \ and\ \bibinfo
  {author} {\bibfnamefont {J.}~\bibnamefont {Oswald}},\ }\href@noop {}
  {\bibfield  {journal} {\bibinfo  {journal} {Journal of Applied Physics}\
  }\textbf {\bibinfo {volume} {102}},\ \bibinfo {pages} {083536} (\bibinfo
  {year} {2007})}\BibitemShut {NoStop}%
\bibitem [{\citenamefont {Klapetek}\ \emph {et~al.}(2011)\citenamefont
  {Klapetek}, \citenamefont {Ne{\v{c}}as}, \citenamefont {Campbellov{\'a}},
  \citenamefont {Yacoot},\ and\ \citenamefont
  {Koenders}}]{klapetek2011methods}%
  \BibitemOpen
  \bibfield  {author} {\bibinfo {author} {\bibfnamefont {P.}~\bibnamefont
  {Klapetek}}, \bibinfo {author} {\bibfnamefont {D.}~\bibnamefont
  {Ne{\v{c}}as}}, \bibinfo {author} {\bibfnamefont {A.}~\bibnamefont
  {Campbellov{\'a}}}, \bibinfo {author} {\bibfnamefont {A.}~\bibnamefont
  {Yacoot}}, \ and\ \bibinfo {author} {\bibfnamefont {L.}~\bibnamefont
  {Koenders}},\ }\href@noop {} {\bibfield  {journal} {\bibinfo  {journal}
  {Measurement Science and Technology}\ }\textbf {\bibinfo {volume} {22}},\
  \bibinfo {pages} {025501} (\bibinfo {year} {2011})}\BibitemShut {NoStop}%
\bibitem [{\citenamefont {Kriegner}\ \emph {et~al.}(2013)\citenamefont
  {Kriegner}, \citenamefont {Wintersberger},\ and\ \citenamefont
  {Stangl}}]{kriegner2013xrayutilities}%
  \BibitemOpen
  \bibfield  {author} {\bibinfo {author} {\bibfnamefont {D.}~\bibnamefont
  {Kriegner}}, \bibinfo {author} {\bibfnamefont {E.}~\bibnamefont
  {Wintersberger}}, \ and\ \bibinfo {author} {\bibfnamefont {J.}~\bibnamefont
  {Stangl}},\ }\href@noop {} {\bibfield  {journal} {\bibinfo  {journal}
  {Journal of applied crystallography}\ }\textbf {\bibinfo {volume} {46}},\
  \bibinfo {pages} {1162} (\bibinfo {year} {2013})}\BibitemShut {NoStop}%
\bibitem [{\citenamefont {Holy}\ \emph {et~al.}(1999)\citenamefont {Holy},
  \citenamefont {Baumbach},\ and\ \citenamefont {Pietsch}}]{holy1999high}%
  \BibitemOpen
  \bibfield  {author} {\bibinfo {author} {\bibfnamefont {V.}~\bibnamefont
  {Holy}}, \bibinfo {author} {\bibfnamefont {T.}~\bibnamefont {Baumbach}}, \
  and\ \bibinfo {author} {\bibfnamefont {U.}~\bibnamefont {Pietsch}},\
  }\href@noop {} {\emph {\bibinfo {title} {High-resolution X-ray scattering
  from thin films and multilayers}}}\ (\bibinfo  {publisher} {Springer},\
  \bibinfo {year} {1999})\BibitemShut {NoStop}%
\bibitem [{\citenamefont {Momma}\ and\ \citenamefont
  {Izumi}(2011)}]{momma2011vesta}%
  \BibitemOpen
  \bibfield  {author} {\bibinfo {author} {\bibfnamefont {K.}~\bibnamefont
  {Momma}}\ and\ \bibinfo {author} {\bibfnamefont {F.}~\bibnamefont {Izumi}},\
  }\href@noop {} {\bibfield  {journal} {\bibinfo  {journal} {Journal of applied
  crystallography}\ }\textbf {\bibinfo {volume} {44}},\ \bibinfo {pages} {1272}
  (\bibinfo {year} {2011})}\BibitemShut {NoStop}%
\bibitem [{com()}]{comsol}%
  \BibitemOpen
  \href@noop {} {\bibinfo  {journal} {COMSOL Multiphysics v. 5.4,
  www.comsol.com, Stockholm, Sweden}\ }\BibitemShut {NoStop}%
\end{thebibliography}%

\end{document}

\title{Supplementary material: Molecular beam epitaxy of CuMnAs}

\pacs{}
% \keywords{CuMnAs, Antiferromagnetism, ....}
\maketitle

%%%%%%%%%%%%%%% Methods %%%%%%%%%%%%%%

\section{Methods}
\paragraph{Substrate and buffer layers} The CuMnAs layers presented in this study were grown on undoped (001) GaP, GaAs and Si substrates. Oxide layer on the GaP and GaAs substrates was desorbed at temperature of 650$^{\circ}$C, and 50 - 100 nm thick buffer layers of GaP or GaAs were grown at 550$^{\circ}$C and 570$^{\circ}$C, respectively. The native oxide on the Si substrate was removed by etching in 3\% HF solution for 45 s immediately before loading into the MBE system. The Si substrate was then heated up to 800$^{\circ}$C and degassed for 1 hour (maximum achievable temperature in our Veeco Gen II system). In this case, the growth of CuMnAs followed directly onto the substrate without any buffer layer.   
%%%%%%%%%%%% Fig. S1 %%%%%%%%%%%%%%%%%%%%%
\begin{figure*}[htbp!]
\vspace{0.2cm}
\includegraphics[scale=0.25]{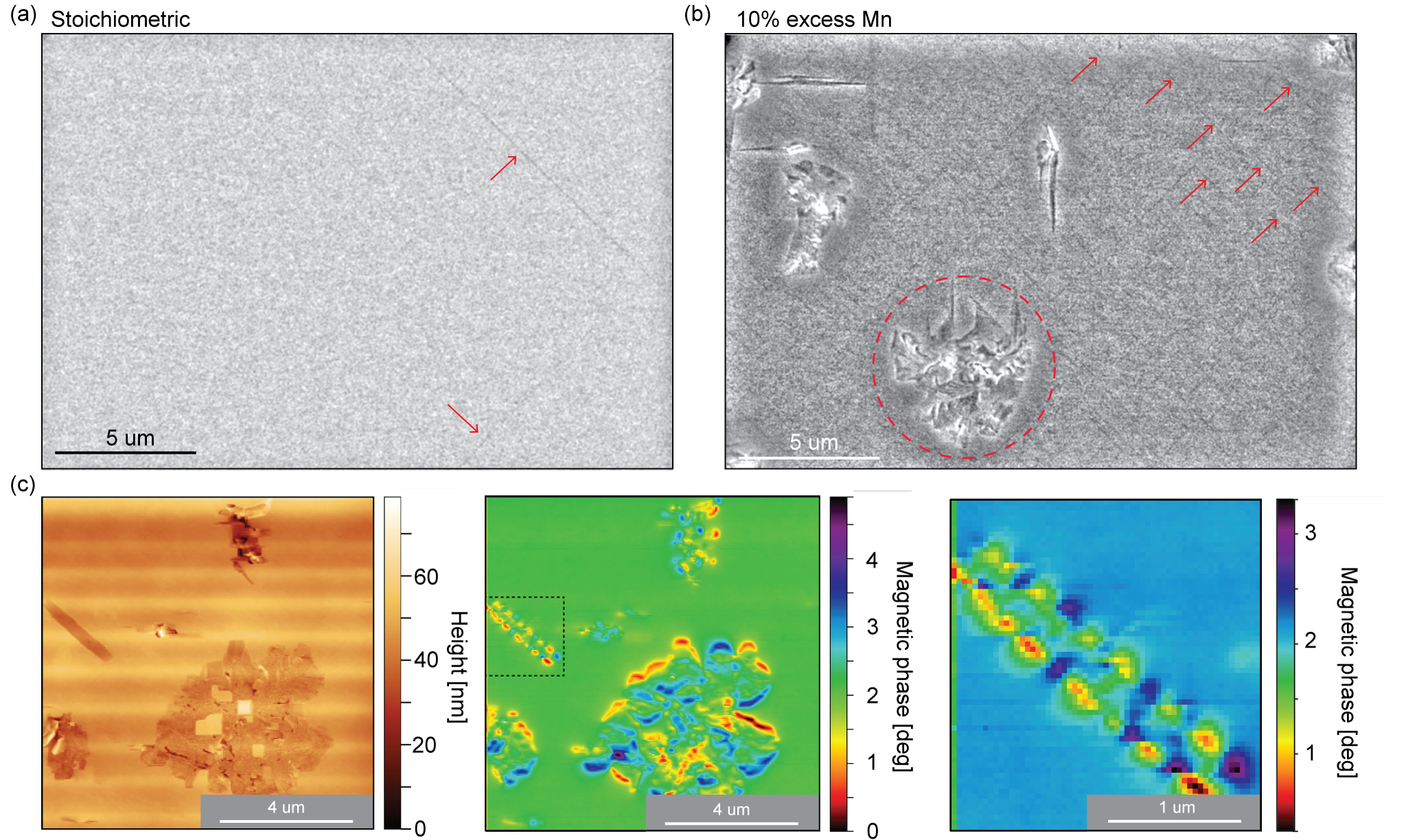}
\vspace{0.2cm}
\caption{SEM micrographs of an (a) a 1:1 Cu:Mn and (b) a Mn-rich CuMnAs samples. The red arrows highlight locations of surface line defects and illustrate their density. The red dashed area highlights an inclusion on the Mn-rich sample. (c) AFM micrograph of the CuMnAs surface with similar inclusions (left panel) and the corresponding magnetic force microscopy image highlighting the magnetic domains (center panel). Detailed scan of highlighted area is shown in the right panel.}
\label{S1}
\end{figure*}
\paragraph{Growth of CuMnAs} After the growth of GaP or GaAs buffer layers (or after thermally treating the Si substrate) a fine adjustment of the beam-equivalent pressures (BEP) of Cu and Mn fluxes were performed. In our case the Cu:Mn BEP ratio of 0.53 was found by the ex-situ EDX and EMPA techniques to correspond to 1:1 stoichiometry in the resulting CuMnAs layer. We estimate that the relative long-term accuracy of the BEP-based flux adjustment is approximately $\pm$ 3\%, which explains the observed scatter in the parameters of the grown layers. After the flux adjustment the substrate temperature was stabilized at the desired growth temperature, measured by the band-edge spectrometry in all cases. The samples were grown betwen 210$^{\circ}$C and 220$^{\circ}$C, if not specified otherwise in the main text. After initiation of the CuMnAs growth the substrate temperature overshoots by $\sim$ 10$^{\circ}$C due to the increasing heat absorption in the growing layer \citen{novak2007substrate}, and stabilizes only within 5 minutes. The CuMnAs was grown at a typical growth rate of 0.021 unit-cell/s.
\paragraph{Al Cap deposition} After CuMnAs growth, the samples were cooled down to $\sim$ 0$^{\circ}$C and 3 nm of Al were deposited with rate of 0.036 ML/s. This results in deposition of slightly granular, but homogeneous layer, that covers the whole sample surface after oxidation at ambient environment. 
\paragraph{AFM measurements} The topography of the samples was characterized on Bruker Dimension Icon Atomic Force Microscope in PeakForce mode. The silicon BudgetSensors Multi75Al-G cantilevers (k = 3 N/m, $f$ = 75 kHz, normal tip radius = 8nm) were used. All samples were measured at the same settings of AFM (scan velocity was 1 $\mu$m/s, PeakForce setpoint $\sim$ 1 nN and PeakForce frequency 1 kHZ). The collected data were processed in the Gwydion software \citen{klapetek2011methods}, and the extracted RMS values correspond to RMS roughness Sq - typically from 2x2 $\mu$m or larger window.  
\paragraph{X-ray analysis} The structure of the samples was analysed by high-resolution X-ray diffractometer equipped with PhotonMax high-flux 9 kW Cu rotating anode and Ge(002x2) monochromator providing a monochromatic CuK$_{\alpha_{1}}$ beam, while using the D/teX Ultra 250 detector. The data were processed using xrayutilities Python package \citen{kriegner2013xrayutilities}. The average size of the crystallographic grains was extracted from reciprocal space maps measurements, as shown in the main text, by comparison with model calculations. For this purpose, model calculation of the diffraction signal of the mosaic films was performed \citen{holy1999high} for the 002 and 105 Bragg peaks. In the used phenomelogical model one assumes that the film consists of randomly rotated mosaic block with averaged lateral and vertical size. The random rotation of these blocks with respect to each other is characterized by the root mean square misorientation. In our case, where Laue thickness fringes can be observed along the $Q_Z$ direction, the vertical block size equivalent to the film thickness can be used. This leaves the lateral block size and root mean square misorientation of the mosaic block as adjustable parameters.
\paragraph{SEM characterization} The samples were characterized in Raith E-line SEM. The surface lines associated with the defects were typically visible above 10k magnification, at 10kV with 20 $\mu$m aperture and the in-beam SE detector.
\paragraph{Lamellae preparation and STEM characterization} The cleaved parts of the samples were adjusted onto standard pin stubs by means of silver lacquer and coated with 10 - 15 nm of carbon using Leica ACE600
coater. The scanning transmission electron microscopy (STEM) lamellae were then fabricated in FEI Helios 660 G3
FIB/SEM following the commonly used protocol, utilizing electron and
ion beam deposited tungsten as protective cap. Final polishing was done at 2 kV at 25 pA. The lamellae were investigated by High resolution (scanning) Transmission Electron Microscope FEI Titan Themis 60-300 cubed at 300 kV acceleration voltage and by FEI probe corrected Analytical Titan at 300 kV. 
\paragraph{Crystal modeling} The models of the crystal structures were done in free crystallographic software Vesta \citen{momma2011vesta}.
\paragraph{Conductivity measurements} The conductivity was measured in the van-der-Pauw geometry either on 5x5 mm samples with In contacts soldered in the corners, or on lithographically patterned cross-shaped samples with bonded contacts. The anisotropy in conductivity was evaluated from the measured values using finite element method simulation of the given sample geometry in COMSOL Multiphysics \citen{comsol}.
\paragraph{Device fabrication} The as grown samples were patterned by standard optical lithography, where 5$\%$ C$_{4}$H$_{13}$NO was used to remove the Al/AlO$_{x}$ cap and CuMnAs was etched with 4:1:2  C$_{4}$H$_{6}$O$_{6}$(5\%):H$_{2}$O$_{2}$(5\%):H$_{2}$SO$_{4}$(10\%) at the etch rate of 50 nm per 30 s.
\paragraph{Electrical switching of CuMnAs} We note that the values for maximal transversal resistivity change were averaged from series of 10 pulses.

\section{S1: Surface line defects and magnetic inclusions}
The surface line defects associated with the micro-twin defects are also visible with SEM. The surface of stoichiometric CuMnAs sample and sample with 10\% excess Mn are shown In Fig. \ref{S1}. There are just 2 defects visible in the 15 x 25 $\mu$m area of the image of the surface of the 1:1 Cu:Mn stoichiometric sample, whereas on the Mn rich sample the defects form a mosaic like structure with spacing bellow 1 $\mu$m.   

In addition, multiple magnetic inclusions are visible at the surface of the Mn rich sample, with spacing on the order of $\mu$m. Their nature is confirmed in Fig. \ref{S1} (c) by magnetic force microscopy. 

\section{S2: Identification of micro-twin tilt in AFM}
\begin{figure}[hb!]
\vspace{0.2cm}
\includegraphics[scale=0.25]{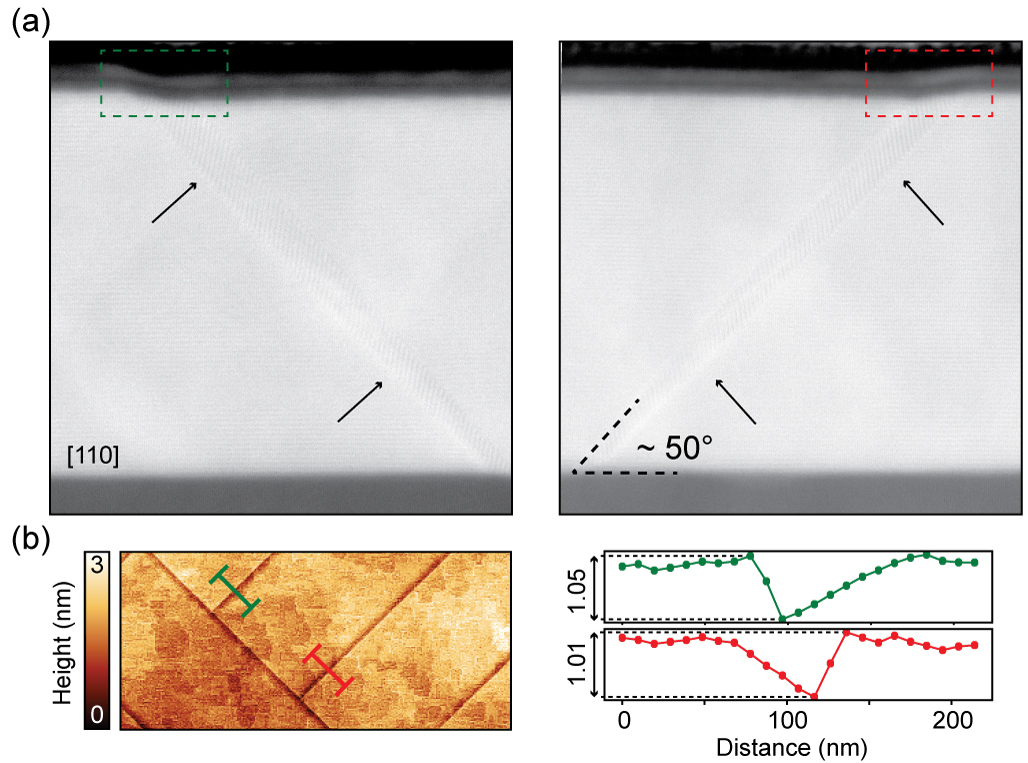}
\vspace{-0.4cm}
\caption{ (a) HAADF-STEM micrographs of two micro-twin defects propagating through the 50 nm of CuMnAs with oposing tilt. (b) AFM scan of the surface of the same sample (different area), where the line cuts are related to the surface steps highlighted in (a).} 
\label{S2}
\end{figure}

As mentioned in the main text, it is possible to identify the tilt of the micro-twin defects within the CuMnAs thin films by AFM. An example is shown in Fig. \ref{S2}. 
In (a), STEM micrographs of two defects with opposite tilt induce a different step onto the surface of the sample. In (b), an AFM scan on the sample (different area), shows that this step is reliably measurable and can be used to identify the angle under which the defect propagates throughout the film. 

%%%%%%%%%%%% Fig. S2 %%%%%%%%%%%%%%%%%%%%%

\section{S3: Strain induced by the micro-twin defects}
\begin{figure}[hb!]
\vspace{0.2cm}
\includegraphics[scale=0.25]{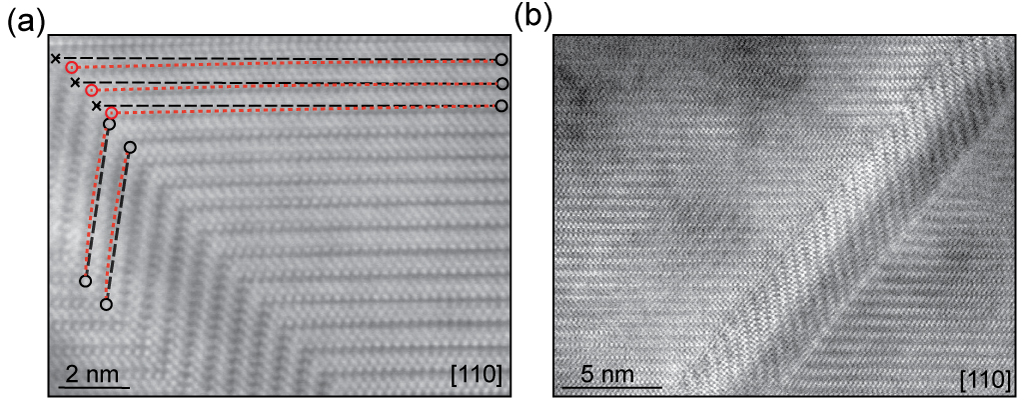}
\vspace{-0.4cm}
\caption{ (a) HAADF-STEM micrograph of a micro-twin defect in CuMnAs thin film. The black lines (guides for the eye) indicate expected position of the atomic rows without strain and the red lines indicate the real position of the atoms. (b) Bright field STEM micrograph of a similar twin defect.} 
\label{S3}
\end{figure}

It is shown in the main text, that the presence of the micro-twin defects is related to tilt visible in the reciprocal space maps. This tilt could be related to strain induced into the CuMnAs layer by presence of the defects, as shown in Fig. \ref{S3}. In (a) it is apparent that the CuMnAs layers in the close proximity of the defect do bend towards the micro-twin boundaries. Also, a strong symmetrical contrast is visible within and around the defect in the bright field STEM image in (b), which is often associated to strain or sample thickness. The first explanation is more probable since it correlates to the XRD measurements and we have observed this trend over multiple defects in multiple samples. The polarity of the contrast is also dependent of the tilt of the defect within the layer.

%%%%%%%%%%%% Fig. S3 %%%%%%%%%%%%%%%%%%%%%

%%%%%%%%%%%% Fig. S4 %%%%%%%%%%%%%%%%%%%%%
\begin{figure*}[h]
\vspace{0.2cm}
\includegraphics[scale=0.25]{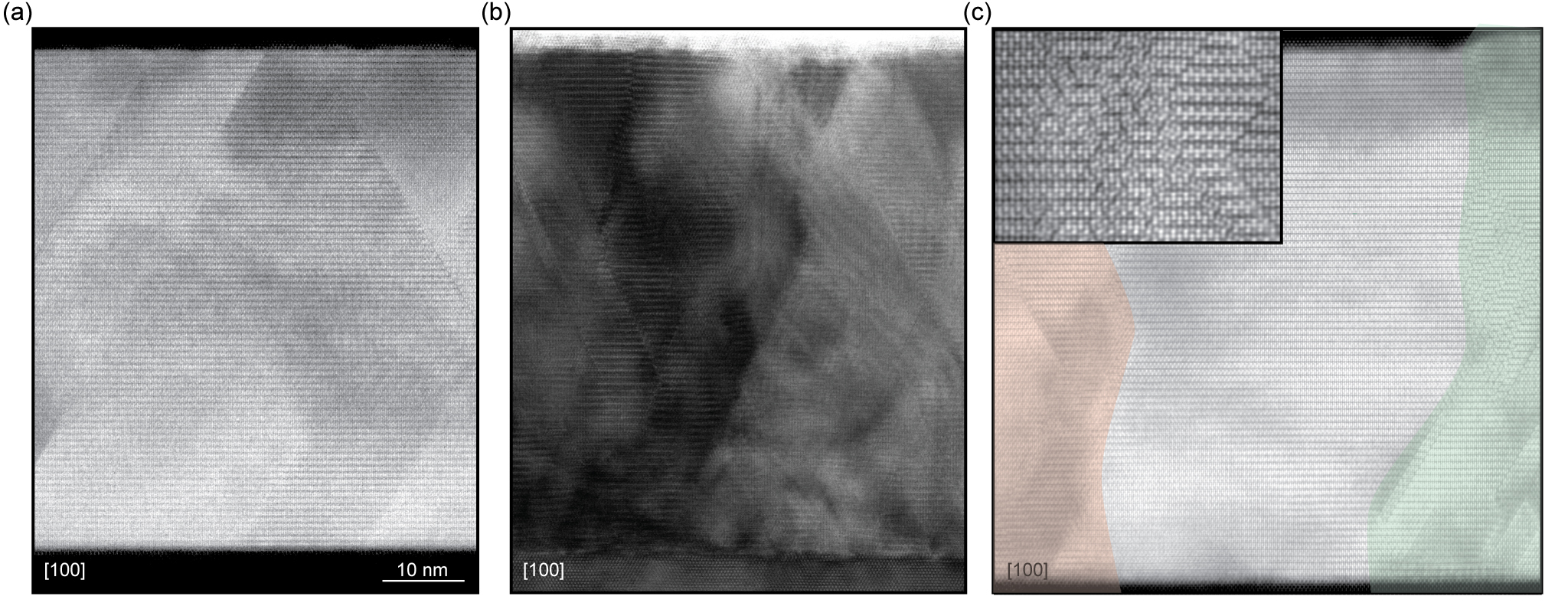}
\vspace{-0.4cm}
\caption{ (a) HAADF-STEM micrograph of CuMnAs grown on GaP. (b) Bright field STEM micrograph of CuMnAs grown on GaAs. (c) HAADF-STEM micrograph of CuMnAs grown on Si. The inset shows zoom-in of part of the area overlayed with green. The red and green overlays show the regions where the otherwise pristine crystal is disturbed from the interface to the top of the film.} 
\label{S4}
\end{figure*} 

%%%%%%%%%%%% Fig. S5 %%%%%%%%%%%%%%%%%%%%%
\begin{figure*}[hb!]
\vspace{0.2cm}
\includegraphics[scale=0.25]{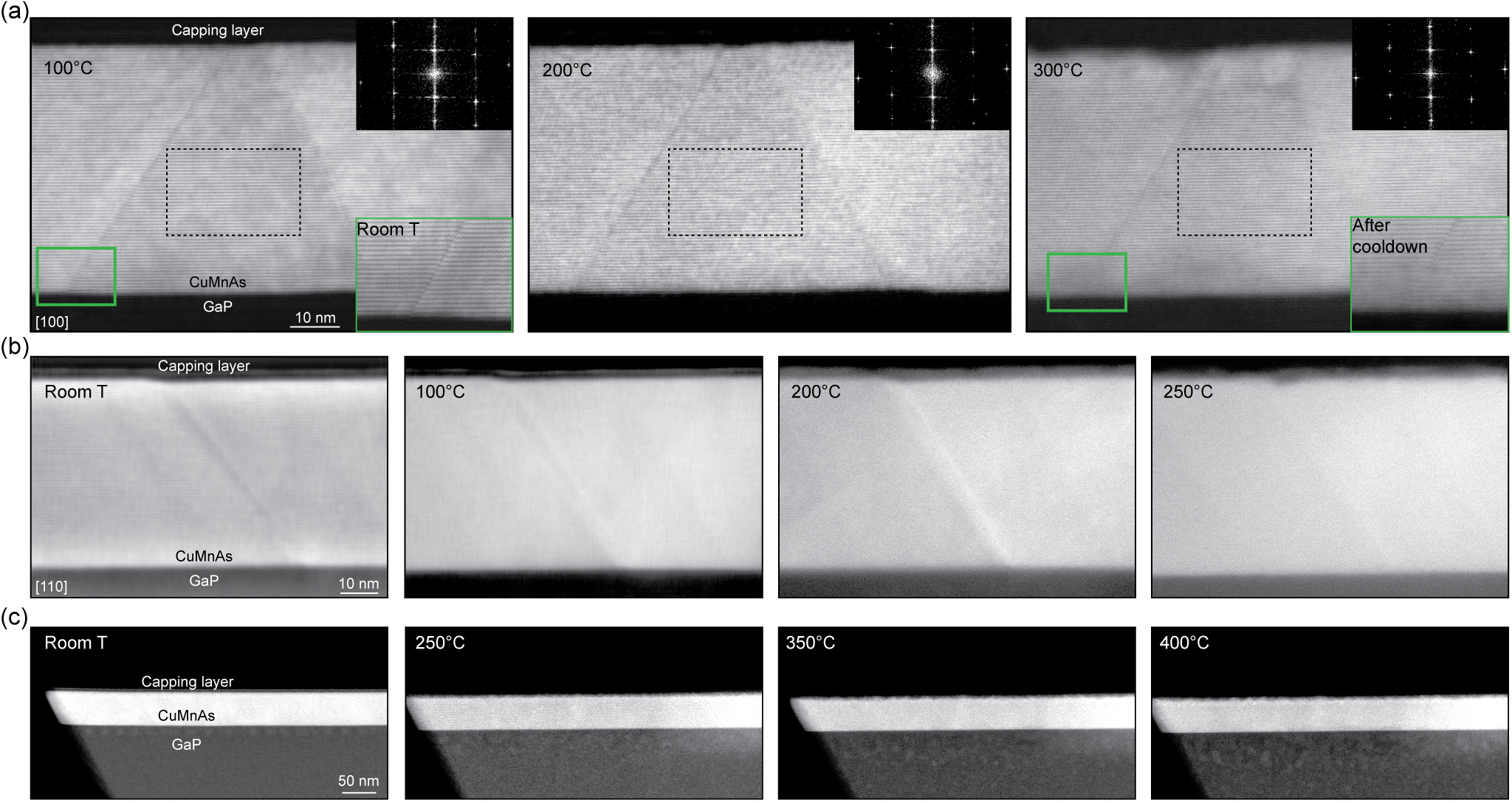}
\vspace{-0.4cm}
\caption{(a) HAADF-STEM micrographs of 50 nm thick CuMnAs layer grown on GaP imaged from the [100] direction at different temperatures, with an anti-phase boundary in the center of the structure. The upper inset shows FFT spectra of the area highlighted by dashed lines. The lower inset in the left panel shows details of the bottom interface (highlighted by green) with an anti-phase boundary defect at room temperature. The lower inset in the right panel shows the same area after cooling from 350$^{\circ}$C to room T. (b) Similar series for the same material but imaged from the [110] direction. A micro-twin defects is visible in the center of the crystal. (c) Zoomed-out micrographs from heating series of the same lamella.} 
\label{S5}
\end{figure*}

%%%%%%%%%%%% Fig. S6 %%%%%%%%%%%%%%%%%%%%%
\begin{figure}[hb!]
\vspace{0.2cm}
\includegraphics[scale=0.25]{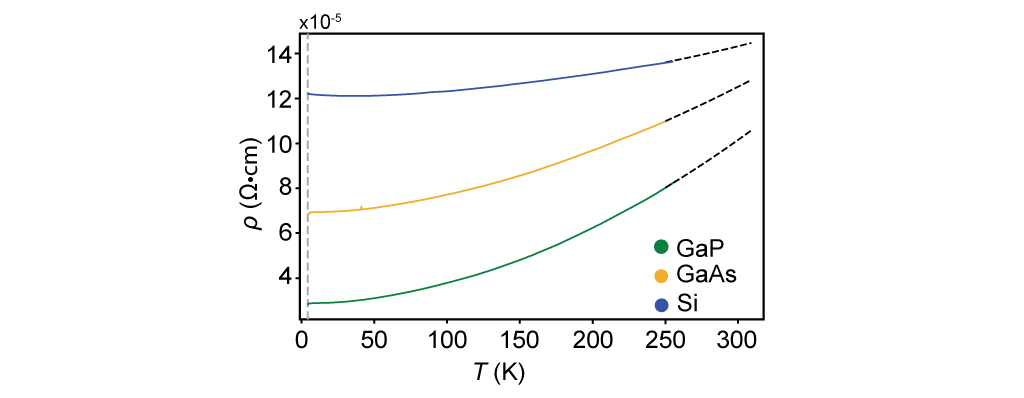}
\vspace{-0.4cm}
\caption{Dependency of resistivity on $T$ for 50 nm thick CuMnAs films grown on GaP (green), GaAs (yellow) and Si (blue). The dashed line are extension of the trends to 300 K.} 
\label{S6}
\end{figure}

\section{S4: Atomic structure of C\lowercase{u}M\lowercase{n}A\lowercase{s} grown on different substrates}
%C\MakeLowercase{u}M\MakeLowercase{n}A\MakeLowercase{s}
As described in the main text, the structure of CuMnAs varies based on the substrate used. On GaP the structure is dominated by the slip dislocations, shown in HAADF-STEM micrograph in Fig. \ref{S4} (a). On GaAs the CuMnAs is significantly strained as apparent in bright field STEM micrograph Fig. \ref{S4} (b). The whole lattice is bending and forms segments that show different contrast in bright field STEM images. The segments are separated by sharp lines following the same direction as are the slip dislocations on GaP. Interestingly, the slip dislocations are often not visible in corresponding HAADF STEM images. Therefore, the phase contrast may not stem from the same type of defects, even though it has the boundary along the same crystal plane. For growth on Si, the layers are disturbed from the bottom interface to the top of the layer. This is likely due to incoherently compensated steps on the surface of the substrate, which result in incoherent crystalline clusters, as shown in HAADF-STEM micrograph in Fig. \ref{S4} (c). Such clusters are often terminated by steps with height up to 20 nm at the top interface and enclose segments of pristine CuMnAs crystals.

\section{S5: Effect of heating on C\lowercase{u}M\lowercase{n}A\lowercase{s}}
%C\lowercase{u}M\lowercase{n}A\lowercase{s}
 
The effect of long term heating on the CuMnAs crystal structure was carried out by heating $\sim$ 80 nm thick lamellae while imaging by scanning STEM. The experiment was carried on 50 thick tetragonal CuMnAs grown on GaP substrates. The result of heating a lamella cut along the [100] direction in CuMnAs is shown in Fig. \ref{S5} (a). The temperature was ramped from room temperature up to 350$^{\circ}$C over 1 hour and 45 minutes and then cooled down back to room T. The Fourier transform spectra in the upper insets clearly do not change with T and show that dominant part of the bulk tetragonal crystal remains unaffected up to 300$^{\circ}$C. Yet, after heating from 200$^{\circ}$C to 300$^{\circ}$C the upper surface of the crystal starts to slowly deteriorate with time. This happens very slowly (order of tens of minutes) at 200$^{\circ}$C and progresses faster with increased temperature. It seems that mainly the capping layer is affected at 200$^{\circ}$C and approximately first 5 MLs were localy affected after staying at 250$^{\circ}$C for approximately 10 minutes. Also, there is a visible recrystallization at the bottom interface, measured for temperatures above 300$^{\circ}$C, which locally affected up to the first 10 MLs and partially erased the anti-phase boundary. This recrystallization remained stable at least for 1 hour after cooldown back to room T, as shown in the lower insets of Fig. \ref{S5} (a).
	
Similar trends are shown in Fig. \ref{S5} (b) for a lamella cut along the [110] direction of CuMnAs. Again, the bulk crystal seems to be unchanged, but the upper interface deteriorates above 200$^{\circ}$C. It is apparent that the micro-twin defect visible in the images remains unchanged during the heating process.
	
A lower magnification overview of a part of similar lamella is shown in Fig. \ref{S5} (c). Again, the same trend for deterioration of the upper interface was apparent above 200$^{\circ}$C. Apart from degradation of the interfaces, there was no clear sign of phase change, melting or other type of damage induced into the bulk of the crystal. We note that during these experiments, only an 80 nm thick slab of the material was exposed to elevated temperatures for time spans longer than 1 hour. This clearly shows that the material retains the same crystal structure at temperatures up to above 200$^{\circ}$C at the interfaces and above 300$^{\circ}$C in the bulk. These temperatures also corresponds to the growth temperature window used in this study.

\section{S6: Temperature dependence of resistance}
As referred to in the main text, there is a difference in the temperature dependency of CuMnAs grown on GaP, GaAs and Si substrates. An example is shown in Fig. \ref{S6} for all the three materials.

\bibliography{supplement}